\documentclass[twocolumn,english,aps,prb,superscriptaddress,bibnotes,amsmath,amssymb,floatfix]{revtex4-1}
\usepackage[colorlinks=true,citecolor=blue,linkcolor=magenta]{hyperref}
\usepackage[utf8]{inputenc}
\usepackage[english]{babel}
\usepackage[T1]{fontenc}
\usepackage{textcomp}
\usepackage[markup=nocolor, authormarkupposition=left]{changes} 
\usepackage{soul}
\usepackage{url}
\usepackage{makecell}
\setaddedmarkup{{\color{red}#1}}
\usepackage{graphicx}
\usepackage{epstopdf}
\graphicspath{{./Figures/}}
\makeatletter
 
\newcommand{\Rmnum}[1]{\expandafter\@slowromancap\romannumeral #1@} 
\makeatother
\usepackage{multirow}

\newcommand{\ket}[1]{\left|#1\right\rangle}
\newcommand{\bra}[1]{\left\langle#1\right|}

\begin{document}
\title{An ultralow-loss integrated photonic platform for\\ discrete-variable quantum information processing}

\author{Ruiyang Chen}
\thanks{These authors contributed equally.}
\affiliation{International Quantum Academy and Shenzhen Futian SUSTech Institute for Quantum Technology and Engineering, Shenzhen 518048, China}
\affiliation{School of Physical Sciences and Hefei National Laboratory, University of Science and Technology of China, Hefei 230026, China}

\author{Zeying Zhong}
\thanks{These authors contributed equally.}
\affiliation{International Quantum Academy and Shenzhen Futian SUSTech Institute for Quantum Technology and Engineering, Shenzhen 518048, China}
\affiliation{Southern University of Science and Technology, Shenzhen 518055, China}

\author{Sanli Huang}
\affiliation{International Quantum Academy and Shenzhen Futian SUSTech Institute for Quantum Technology and Engineering, Shenzhen 518048, China}
\affiliation{School of Physical Sciences and Hefei National Laboratory, University of Science and Technology of China, Hefei 230026, China}

\author{Sicheng Zeng}
\affiliation{International Quantum Academy and Shenzhen Futian SUSTech Institute for Quantum Technology and Engineering, Shenzhen 518048, China}
\affiliation{Southern University of Science and Technology, Shenzhen 518055, China}

\author{Zhenyuan Shang}
\affiliation{International Quantum Academy and Shenzhen Futian SUSTech Institute for Quantum Technology and Engineering, Shenzhen 518048, China}
\affiliation{Southern University of Science and Technology, Shenzhen 518055, China}

\author{Yue Hu}
\affiliation{International Quantum Academy and Shenzhen Futian SUSTech Institute for Quantum Technology and Engineering, Shenzhen 518048, China}
\affiliation{Southern University of Science and Technology, Shenzhen 518055, China}

\author{Zhen Chen}
\affiliation{International Quantum Academy and Shenzhen Futian SUSTech Institute for Quantum Technology and Engineering, Shenzhen 518048, China}
\affiliation{School of Physical Sciences and Hefei National Laboratory, University of Science and Technology of China, Hefei 230026, China}

\author{Yuan Chen}
\affiliation{International Quantum Academy and Shenzhen Futian SUSTech Institute for Quantum Technology and Engineering, Shenzhen 518048, China}

\author{Shuyi Li}
\affiliation{International Quantum Academy and Shenzhen Futian SUSTech Institute for Quantum Technology and Engineering, Shenzhen 518048, China}

\author{Xue Bai}
\affiliation{International Quantum Academy and Shenzhen Futian SUSTech Institute for Quantum Technology and Engineering, Shenzhen 518048, China}
\affiliation{Qaleido Photonics, Shenzhen 518048, China}

\author{Yi-Han Luo}
\email{luoyh@iqasz.cn}
\affiliation{International Quantum Academy and Shenzhen Futian SUSTech Institute for Quantum Technology and Engineering, Shenzhen 518048, China}

\author{Junqiu Liu}
\email{liujq@iqasz.cn}
\affiliation{International Quantum Academy and Shenzhen Futian SUSTech Institute for Quantum Technology and Engineering, Shenzhen 518048, China}
\affiliation{School of Physical Sciences and Hefei National Laboratory, University of Science and Technology of China, Hefei 230026, China}

\maketitle

\noindent \textbf{Photonic integrated circuits offer a scalable and robust route toward quantum information technologies by consolidating photon sources and linear optical networks onto compact, wafer-manufacturable chips.
Although silicon photonics has enabled diverse discrete-variable quantum breakthroughs---spanning multiphoton entanglement, quantum networking, and photonic qubit fusion for quantum computing---scaling these platforms beyond proof-of-principle demonstrations remains constrained by a critical system-level bottleneck. 
Optical loss compounds rapidly across photon generation, routing, and state analysis, causing multiphoton generation probabilities to plummet exponentially as circuit depth and complexity grow. 
Here we overcome this rate--loss barrier by demonstrating a monolithic, ultralow-loss silicon nitride (Si$_3$N$_4$) integrated photonic platform engineered for high-performance discrete-variable quantum information processing.
Our architecture seamlessly integrates narrowband photon-pair sources with low-loss qubit-fusion circuits and reconfigurable state-analysis interferometers. 
The on-chip sources prepare Einstein--Podolsky--Rosen (EPR) states with a fidelity of 0.9875(3) and exhibit near-unity photon indistinguishability, yielding a heralded Hong--Ou--Mandel interference visibility of 0.990(6). 
By executing on-chip fusion of two EPR states, we synthesize and characterize four-photon Greenberger--Horne--Zeilinger states with a record fidelity of 0.943(8) and a fourfold count rate of 27 Hz---more than two orders of magnitude higher than previous silicon-photonic implementations.
Combined with standard CMOS-compatible fabrication on 150-mm-diameter wafers, these results establish ultralow-loss Si$_3$N$_4$ integrated photonics as a manufacturable platform for deployable, large-scale quantum information processors.
}

Photons constitute a versatile and inherently robust platform for quantum information processing (QIP), enabling high-fidelity state generation, coherent manipulation, and efficient detection \cite{Pan:12, Flamini:18}.
These capabilities have underpinned milestone demonstrations of quantum computational advantage \cite{Zhong:20, Madsen:22, Liu:26}, quantum networking \cite{Yin:17, Ren:17, Chen:21}, and precision metrology \cite{Nagata:07, Giovannetti:11, LiuLZ:21}.
A central challenge now lies in translating these laboratory breakthroughs into deployable and scalable systems without compromising state-of-the-art performance. 
Photonic integrated circuits (PICs) offer a compelling path forward by integrating sources, linear networks, and detectors onto compact, phase-stable chips compatible with wafer-scale volume manufacturing \cite{WangJW:20a, Pelucchi:22}.

Within integrated architectures, path-encoded discrete-variable (DV) photonic qubits provide an elegant framework, as on-chip waveguides and interferometers directly define the qubit basis while executing complex coherent operations. 
Crucially, optical loss in DV encodings manifests as photon erasure rather than continuous state degradation \cite{Pankovich:24}, while the probabilistic nature of linear optics can be rendered scalable via heralding and multiplexing \cite{KLM:01, Bartolucci:23}. 
Leveraging these advantages, silicon photonics has driven a succession of pioneering proof-of-principle demonstrations, spanning multiphoton entangled states generation \cite{Kues:17, WangJW:18, Bao:23, ChenL:23}, quantum networking protocols \cite{Llewellyn:20, Zheng:23}, and photonic qubit fusion toward universal quantum computing \cite{Alexander:25}.
However, these scaling efforts have exposed a severe system-level bottleneck: 
a multiphoton operation succeeds only when every photon survives propagation and is detected. 
As the photon number and circuit depth increase, compounding propagation and component losses exponentially suppress the multiphoton coincidence rate. 
Overcoming this rate--loss barrier demands an integrated platform that simultaneously delivers ultralow propagation loss and bright, mutually indistinguishable photon sources.

\begin{figure*}[t!]
\centering
\includegraphics[width=\linewidth]{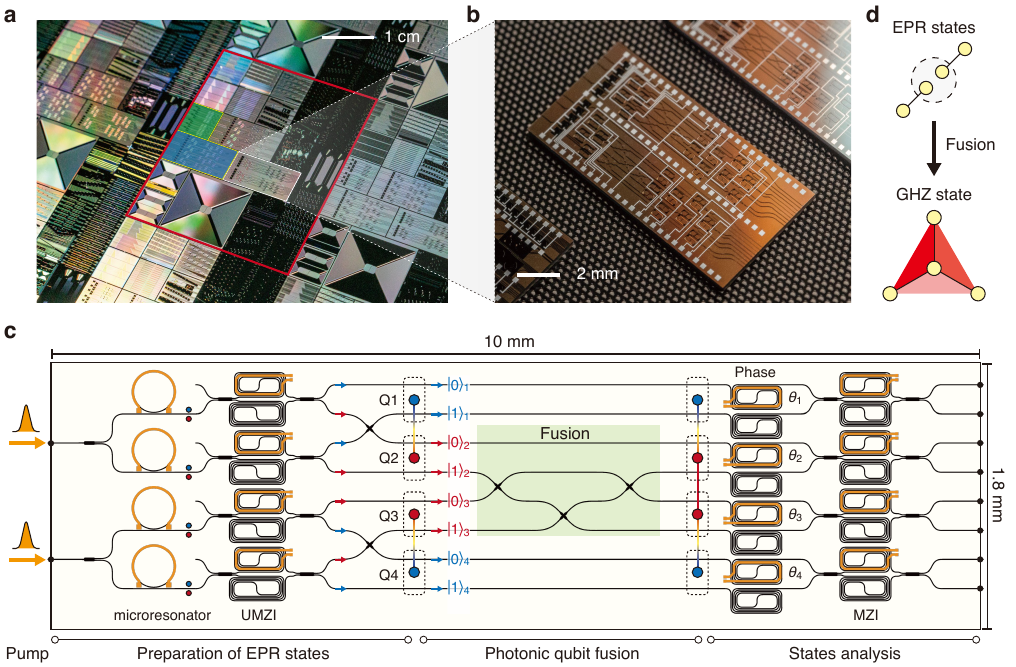}
\caption{
\textbf{Wafer-scale ultralow-loss silicon nitride photonic integrated circuits for discrete-variable quantum information processing.}
\textbf{a}, 
Photograph of a 150-mm-diameter multi-project wafer hosting ultralow-loss Si$_3$N$_4$ PICs fabricated with a foundry-scale CMOS-compatible process. 
The red outline highlights a 30 mm $\times$ 25 mm reticle field containing functionally distinct chips, including the target device for four-photon GHZ state generation (enlarged in \textbf{b}), alongside separate chips dedicated to photon indistinguishability measurements (shaded green) and EPR state characterization (shaded blue).
\textbf{c}, 
Schematic of the 10 mm $\times$ 1.8 mm PIC for synthesizing four-photon GHZ states. 
Pulsed pump lasers are launched from the left chip edge to drive the microresonator-based photon-pair sources. 
Following spectral demultiplexing by UMZIs, a path-exchange network routes the generated photon pairs into two independent, path-encoded EPR states, labeled Q1--Q2 and Q3--Q4. 
On-chip fusion of qubits Q2 and Q3 projects the two EPR states into a four-photon GHZ state via post-selection. 
The four output photonic qubits are then routed to individual state-analysis modules, where reconfigurable MZIs select the measurement basis (switching between the computational and superposition bases), while thermally tuned delay lines independently control the local phases $\theta_{1-4}$.
\textbf{d}, 
Graph-state representation of the four-photon GHZ state generation process.
Two independent EPR states are fused into a four-photon GHZ state, where vertices represent the photonic qubits and the connecting edges represent entangling operations.
}
\label{Fig:1}
\end{figure*}

\begin{figure*}[t!]
\centering
\includegraphics[width=\linewidth]{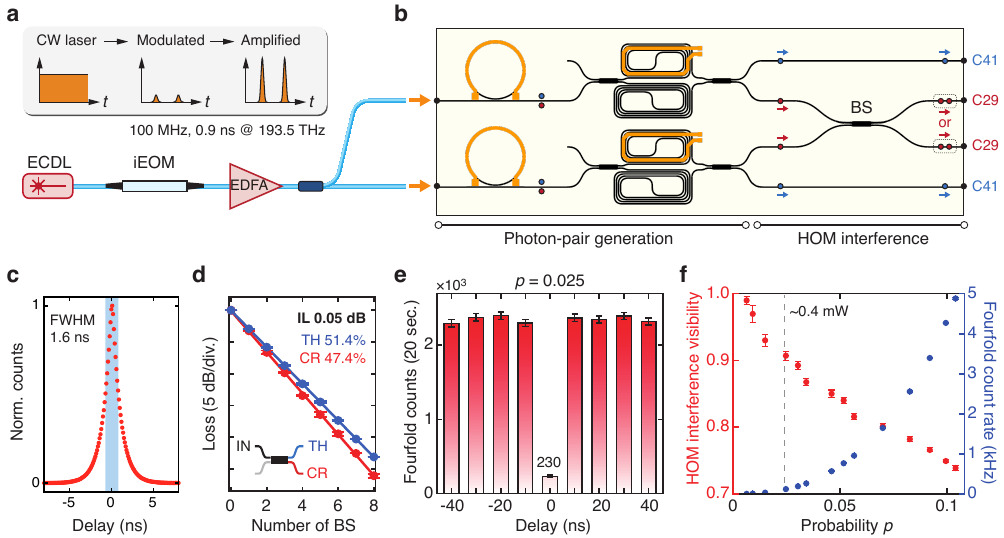}
\caption{
\textbf{Characterization of on-chip photon-pair source indistinguishability.}
\textbf{a}, 
Optical pump preparation. 
CW light from an ECDL is temporally carved using an iEOM and subsequently amplified by an EDFA, to prepare a 100 MHz train of 0.9 ns optical pulses centered at 193.5 THz (channel C35). 
\textbf{b}, 
Schematic of the on-chip HOM interference circuit.
Two independent microresonator sources generate photon pairs in channels C29 and C41. 
UMZIs demultiplex the co-propagating modes. 
The C41 photons are detected off-chip as heralding triggers, while their partner C29 photons are routed to interfere at an on-chip BS.
Bunching of identical C29 photons at the BS suppresses fourfold coincidence events. 
\textbf{c}, 
Photon-pair correlation histogram. 
The measured FWHM of 1.6 ns is consistent with a microresonator resonance linewidth of approximately 160 MHz, validating the narrow linewidth of the sources. 
\textbf{d}, 
Characterization of the multimode interference BS using cascaded test structures. 
A through (TH) transmission of 51.4\% and a cross (CR) transmission of 47.4\% are extracted, corresponding to a low insertion loss of 0.05 dB per device. 
\textbf{e}, 
Fourfold counts versus relative temporal delay. 
Evaluated at a photon-pair generation probability of $p=0.025$, the zero-delay coincidence bin drops to 230 counts over a 20-s integration time, while large delays establish the distinguishable-photon baseline. 
\textbf{f}, 
HOM interference visibility (red, left axis) and large-delay baseline fourfold count rate (blue, right axis) plotted against the photon-pair generation probability $p$. 
The vertical dashed line identifies the $p=0.025$ operating point, which corresponds to an on-chip pump power of approximately 0.4 mW.
}
\label{Fig:2}
\end{figure*}

In this work, we overcome this limitation by establishing an ultralow-loss Si$_3$N$_4$ integrated photonic platform \cite{Liu:21, Ye:23} engineered specifically for high-performance DV-QIP. 
Our architecture monolithically integrates the entire functional stack: Einstein--Podolsky--Rosen (EPR) sources \cite{Einstein:1935, Kwiat:1995}, low-loss qubit-fusion circuits, and reconfigurable state-analysis interferometers. 
The on-chip EPR sources simultaneously exhibit high state fidelity and near-unity photon indistinguishability, which, combined with the exceptionally low propagation loss of the passive circuits, establishes a highly efficient foundation for scalable quantum logic. 
All devices are fabricated using a standard CMOS-compatible foundry process on 150-mm-diameter multi-project wafers (Fig. \ref{Fig:1}a; see Methods for details), directly bridging high-fidelity quantum photonic operations with scalable industrial manufacturing. 

To benchmark the system-level performance of our platform, we demonstrate the generation and analysis of a four-photon Greenberger--Horne--Zeilinger (GHZ) state \cite{GHZ}---a critical resource for quantum communication networks \cite{Xiao:04, Chen:05, Lu:16} and entanglement-enhanced metrology \cite{Nagata:07, Giovannetti:11, LiuLZ:21}.
This execution combines EPR states preparation with on-chip photonic qubit fusion, validating the entire source--fusion--analysis chain of linear optical DV-QIP \cite{KLM:01} on a single, monolithic device (Fig. \ref{Fig:1}b). 
We achieve a four-photon GHZ state fidelity of up to 0.943(8), the highest reported to date in integrated photonics, alongside a fourfold count rate of 27 Hz---more than two orders of magnitude higher than prior silicon-photonic implementations \cite{Llewellyn:20, ChenLZ:24}. 
Crucially, the platform preserves narrow, 100-MHz-level photon linewidths \cite{Chen:2024, Li:25}, ensuring robust indistinguishability over long-distance fiber transmission and enabling efficient interfacing with quantum memories \cite{LiuX:21a, LiuW:26, LuB:26}. 

The layout of the monolithic PIC for four-photon GHZ state generation is illustrated schematically in Fig. \ref{Fig:1}c. 
Pulsed pump lasers launch into the chip edge and propagate through three sequential functional stages: EPR state preparation, photonic qubit fusion, and state analysis. 
In the preparation stage, four microresonator-based sources generate photon pairs via cavity-enhanced spontaneous four-wave mixing (SFWM) \cite{Helt:10}.
Unbalanced Mach--Zehnder interferometers (UMZIs) demultiplex these pairs into distinct spatial modes, which are then routed via path exchange to construct two independent, path-encoded EPR states labeled Q1--Q2 and Q3--Q4. 

Next, the fusion circuit executes a path-exchange operation on photons Q2 and Q3, projecting the two independent EPR states into a four-photon GHZ state via post-selection. 
Finally, each qubit is routed to an independent state-analysis module comprising a thermally tunable optical delay line and a reconfigurable Mach--Zehnder interferometer (MZI). 
The MZIs select the measurement basis---switching between the computational and balanced superposition bases---while the delay lines independently tune the local phases $\theta_{1-4}$. 
This entire circuit sequence maps directly onto the graph-state representation \cite{Hein:04} shown in Fig. \ref{Fig:1}d, where vertices denote the photonic qubits and edges represent entangling operations.
In the following sections, we systematically validate this operational stack by first evaluating the on-chip Hong--Ou--Mandel (HOM) interference that underpins high-fidelity fusion, benchmarking the EPR sources, and finally characterizing the synthesized four-photon GHZ state.

\noindent\textbf{Near-unity photon indistinguishability.}
High-visibility multiphoton interference requires near-unity photon indistinguishability, which dictates the fidelity of qubit-fusion operations.
For a parametric photon-pair source, indistinguishability is typically benchmarked via HOM interference \cite{Hong:87} between independent heralded single photons.
Consequently, the HOM interference visibility serves as a primary metric governing ultimate scaling capacity of DV-QIP architectures.

The experimental configuration for benchmarking source indistinguishability is illustrated in Fig. \ref{Fig:2}a and b. 
Continuous-wave (CW) light from an external-cavity diode laser (ECDL) is temporally carved using an intensity electro-optic modulator (iEOM) and amplified with an erbium-doped fiber amplifier (EDFA) to produce a 100 MHz train of 0.9 ns pulses centered at 193.5 THz (channel C35). 
These pump pulses exhibit a near-transform-limited bandwidth of approximately 500 MHz. 
This pulse train is equally split to pump two independent microresonator-based photon-pair sources. 
Details of the experimental setup are provided in Supplementary Information Note 1. 
Fabricated from 800-nm-thick Si$_3$N$_4$ waveguides engineered for anomalous group-velocity dispersion \cite{Ye:23, Chen:2024} within our CMOS foundry, these microresonators generate photon pairs via cavity-enhanced SFWM \cite{Helt:10} centered at 192.9 THz (channel C29, signal) and 194.1 THz (channel C41, idler). 
Details of the microresonator design and characterization are provided in Methods and Supplementary Information Note 1.
Following generation, on-chip UMZIs demultiplex the co-propagating photon pairs, routing the C41 photons off-chip to single-photon detectors to act as heralds, while directing the partner C29 photons to an on-chip beam splitter (BS). 
The simultaneous arrival of two identical heralded C29 photons leads to HOM bunching at the BS, suppressing fourfold coincidence events.

The photon-pair correlation histogram (Fig. \ref{Fig:2}c) confirms robust pair generation, featuring a full width at half maximum (FWHM) of 1.6 ns. 
This temporal profile is consistent with the microresonator's loaded resonance linewidth of approximately 160 MHz, verifying the narrow linewidth of the generated photons. 
The on-chip BS is implemented via a multimode interferometer (MMI) \cite{Soldano:95}, which we characterize using cascaded test structures (Fig. \ref{Fig:2}d). 
The extracted cross (CR) and through (TH) transmissions are 47.4\% and 51.4\%, respectively, corresponding to an exceptionally low insertion loss of 0.05 dB per device. 
This slight splitting imbalance is accounted for during the extraction of the HOM interference visibility. 
Details on device characterizations and splitting-imbalance corrections are presented in Supplementary Information Notes 1 and 2.

\begin{figure*}[t!]
\centering
\includegraphics[width=\linewidth]{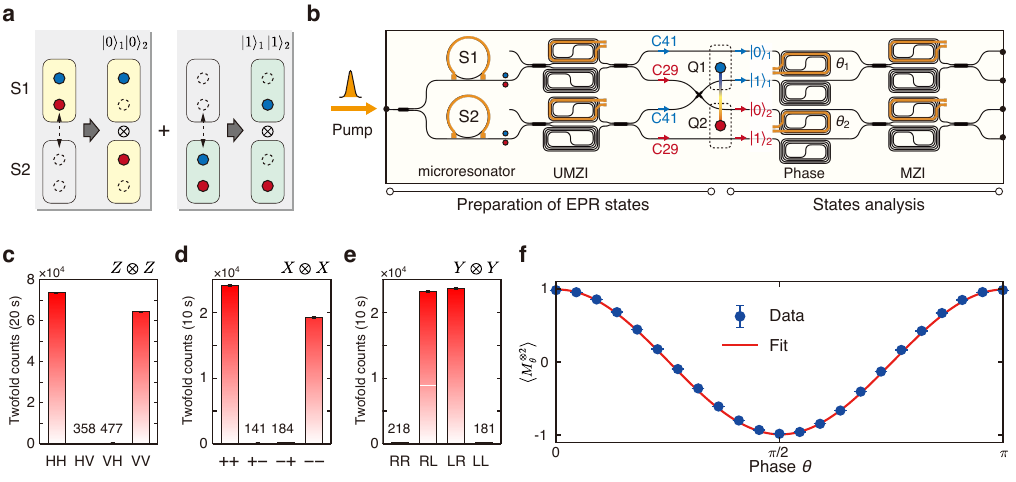}
\caption{
\textbf{Generation and characterization of the path-encoded EPR state. }
\textbf{a}, 
Operational principle of path-encoded EPR state generation. 
Twin photon-pair sources (S1 and S2) each generate photon pairs into dual spatial paths that define the qubit basis. 
By spatially swapping the lower path of S1 with the upper path of S2, photon-pair generation from S1 populates the upper spatial modes to yield $\ket{0}_1\ket{0}_2$, whereas photon-pair generation from S2 populates the lower spatial modes to yield $\ket{1}_1\ket{1}_2$.
Under coherent pumping, these two pathways superpose into an EPR state. 
\textbf{b}, 
Schematic of the characterization circuit. 
Twin microresonators act as sources S1 and S2 driven by a common pulsed pump. 
Following mode demultiplexing via the UMZI, the spatial paths of the C29 and C41 photons are cross-routed to prepare the path-encoded EPR state. 
Each qubit is analyzed using an independent thermo-optic phase shifter and a reconfigurable MZI. 
\textbf{c--e}, 
Twofold counts recorded across the $Z\otimes Z$ (\textbf{c}), $X\otimes X$ (\textbf{d}), and $Y\otimes Y$ (\textbf{e}) bases, accumulated over 20 s, 10 s, and 10 s, respectively. 
The marked minima identify the strongly suppressed outcomes in each basis: the cross-terms $\ket{0}_1\ket{1}_2$ and $\ket{1}_1\ket{0}_2$ for $Z\otimes Z$, opposite-parity outcomes for $X\otimes X$, and same-parity outcomes for $Y\otimes Y$. 
\textbf{f}, 
Two-photon quantum interference fringe. 
The measured expectation value $\langle M_\theta^{\otimes2}\rangle$ is plotted as a function of the local phase $\theta$, where $M_\theta=\cos\theta\,X+\sin\theta\,Y$. 
The fitted $\pi$-period fringe oscillates at exactly twice the single-qubit phase evolution frequency, confirming two-photon coherence in the EPR state.
}
\label{Fig:3}
\end{figure*}

To evaluate the HOM interference visibility as a function of on-chip photon flux, we define $p$ as the probability of generating a single photon pair per pump pulse within an individual source. 
At $p=0.025$ (corresponding to an on-chip pump power of $\sim$0.4 mW), the HOM interference demonstrates a pronounced suppression of fourfold count at zero relative delay, recording 230 counts over 20-s integration (Fig. \ref{Fig:2}e).
Introducing large channel delays eliminates the temporal wavepacket overlap at the BS, destroying the interference and establishing a distinguishable-photon baseline of more than 2,000 counts (see Supplementary Information Note 2).
Mapping the HOM interference visibility against $p$ (Fig. \ref{Fig:2}f) reveals the performance trade-off governed by multi-pair emission. 
At $p=0.006$, the visibility reaches a near-unity value of 0.990(6) at a baseline fourfold count rate of 1.5 Hz. 
Crucially, even at the higher-flux operating point of $p = 0.025$, where multi-pair noise degrades the interference visibility, it remains above 0.90 while increasing the fourfold count rate beyond 100 Hz.

\begin{figure*}[t!]
\centering
\includegraphics[width=\linewidth]{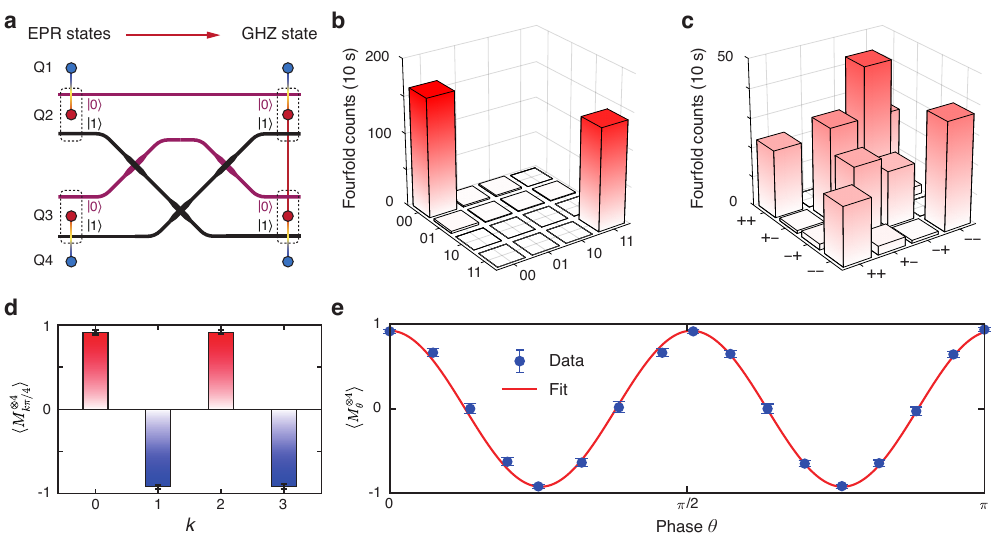}
\caption{
\textbf{Four-photon GHZ state generation and characterization via on-chip photonic qubit fusion.}
\textbf{a}, 
Schematic of four-photon GHZ state generation. 
The integrated fusion circuit swaps the logical $\ket{1}$ modes of the path-encoded qubits Q2 and Q3 originating from two separate EPR states (Q1--Q2 and Q3--Q4). 
Post-selection projects the system into the four-photon GHZ state when exactly one photon is registered in each output qubit channel. 
\textbf{b}, 
Fourfold counts in the computational basis accumulated over a 10-s integration interval, displaying high-contrast state populations concentrated in the $\ket{0}^{\otimes 4}$ and $\ket{1}^{\otimes 4}$ components.
\textbf{c}, 
Fourfold counts in the basis $|\pm\rangle=\left(\ket{0}\pm\ket{1}\right)/\sqrt{2}$ over a 10-s integration interval, illustrating the parity-dependent multiphoton quantum interference.
\textbf{d}, 
Measured expectation values $\langle M_{k\pi/4}^{\otimes 4}\rangle$ for $k=0,1,2,3$, used to extract the state coherence $\langle C\rangle$. 
\textbf{e}, 
Four-photon quantum interference fringe. 
The expectation value $\langle M_{\theta}^{\otimes 4}\rangle$ is tracked as a function of the local phase $\theta$. 
The red curve represents a sinusoidal fit proportional to $\cos 4\theta$, verifying the fourfold phase-oscillation frequency expected for a four-photon GHZ state.
}
\label{Fig:4}
\end{figure*}

\noindent\textbf{Characterization of the path-encoded EPR state.}
EPR states \cite{Einstein:1935} serve as foundational resources for photonic QIP, underpinning both fundamental tests of quantum nonlocality and diverse applications across quantum networking, computing, and metrology.
The principle for generating path-encoded EPR states is outlined in Fig. \ref{Fig:3}a. 
Two identical photon-pair sources, S1 and S2, each generate photon pairs into two distinct spatial paths that define the qubit computational basis, where the upper and lower paths encode $\ket{0}$ and $\ket{1}$, respectively. 
A path exchange swaps the lower path of S1 with the upper path of S2.
Consequently, a photon pair generated in S1 populates the upper spatial paths, yielding the state $\ket{0}_1\ket{0}_2$, whereas a photon pair generated in S2 populates the lower paths, yielding $\ket{1}_1\ket{1}_2$.
Under weak pumping conditions where multi-pair generation is negligible, driving both S1 and S2 with a shared pump laser renders these two terms into a coherent superposition, 
synthesizing the EPR state
\begin{equation}
\ket{\Phi^+}=\frac{1}{\sqrt{2}}\left(\ket{0}_1\ket{0}_2+\ket{1}_1\ket{1}_2\right).
\end{equation}
The physical realization of this scheme is shown in Fig. \ref{Fig:3}b.
Two microresonators serving as S1 and S2 are driven by a common pulsed pump.
Following on-chip UMZI demultiplexing, the paths of the C29 photon from S1 and the C41 photon from S2 are cross-routed, preparing the path-encoded EPR state within the C29--C41 photon pair.

We characterize the generated EPR state by measuring its state fidelity $F$, which is obtained via the Pauli decomposition of the ideal EPR state projector:
\begin{equation}
F=\frac{1}{4}
\left(
1+\langle Z\otimes Z\rangle
+\langle X\otimes X\rangle
-\langle Y\otimes Y\rangle
\right),
\end{equation}
where $X$, $Y$, and $Z$ are the Pauli operators. 
Details of the state fidelity definition are provided in Supplementary Information Note 2. 
The eigenstates corresponding to the $\{+1,-1\}$ eigenvalues for the $Z$, $X$, and $Y$ bases are defined as $\{\ket{0},\ket{1}\}$, $\{\ket{+},\ket{-}\}$, and $\{\ket{R},\ket{L}\}$, respectively.
As illustrated in Fig. \ref{Fig:3}b, these projection measurements are implemented using a thermally tunable delay line acting as a phase shifter, followed by a reconfigurable MZI. 
Configuring the MZI for direct path transmission permits measurement in the $Z$ basis $\{\ket{0},\ket{1}\}$.
When configured as a balanced BS, the MZI projects onto the superposition basis $\left(\ket{0}\pm e^{i\theta}\ket{1}\right)/\sqrt{2}$, where $\theta=\theta_{1,2}$ are set individually by the two integrated phase shifters.
Choosing $\theta=0$ or $\theta=\pi/2$ selects the $X$ or $Y$ measurement basis, respectively.

The twofold counts acquired across the $Z\otimes Z$, $X\otimes X$, and $Y\otimes Y$ bases are presented in Figs. \ref{Fig:3}c--e. 
The high-contrast minima observed in each basis demonstrate strong population correlations and phase-sensitive coherence. 
From these data, we extract a near-unity EPR state fidelity of $F=0.9875(3)$. 
This outstanding fidelity is fundamentally enabled by the high fabrication uniformity of our Si$_3$N$_4$ platform, which ensures precise spectral alignment of the C29 and C41 resonances across the constituent microresonator sources (see Methods). 
We further map the two-photon quantum interference fringes. 
For a single qubit, the observable $M_\theta=\cos\theta\,X+\sin\theta\,Y$ possesses the eigenstates $\left(\ket{0}\pm e^{i\theta}\ket{1}\right)/\sqrt{2}$ with eigenvalues $+1$ and $-1$.
Figure \ref{Fig:3}f plots the measured expectation value $\langle M_\theta^{\otimes2}\rangle$ as a function of $\theta$ varied from $0$ to $\pi$.
The sinusoidal fit exhibits a clear $\pi$-period oscillation---exactly twice the frequency of single-qubit phase evolution---confirming the two-photon coherence of the EPR state.

\begin{table*}[t!]
\caption{
\textbf{Performance comparison of state-of-the-art photonic integrated circuits for discrete-variable quantum information processing.} 
The benchmarked metrics include the HOM interference visibility between independent heralded single photons, 
the corresponding large-delay fourfold count rate (defining the distinguishable-photon baseline), 
the four-photon GHZ state fidelity, 
and its associated fourfold count rate. 
Values that were not explicitly reported or cannot be reliably inferred from the original literature are denoted by ``$-$''. 
The Si$_3$N$_4$ work by Samara \emph{et al.} \cite{Samara:21} is included exclusively for source-level comparison, as their photon-pair sources were integrated on-chip while the subsequent HOM interference was executed using off-chip fiber-optic components. 
Abbreviations: SPDC, spontaneous parametric down-conversion; 
MR, microresonator; 
WG, waveguide; 
TFLN, thin-film lithium niobate;
QD, quantum dot.
}
\begin{ruledtabular}
\begin{tabular}{lccc@{\hspace{0.3cm}}c@{}c@{\hspace{0.3cm}}c@{}c}
\multirow{2}*{Reference} & \multirow{2}*{Platform} & \multirow{2}*{Source type}& Linewidth & HOM interference & Baseline count & Four-photon GHZ & GHZ count \\
 & & & (GHz) & visibility & rate (Hz)& state fidelity & rate (Hz) \\[1.5pt]
\colrule\\[-8pt]
\multirow{2}*{This work}  & \multirow{2}*{Si$_3$N$_4$} & \multirow{2}*{SFWM/MR} & \multirow{2}*{0.16} & \multirow{2}*{0.990(6)} & \multirow{2}*{1.5} & 0.943(8) & 27 \\[1.6pt]
 & & & & & & 0.860(10) &  134 \\[1.6pt]
\colrule\\[-6pt]
Alexander \emph{et al.} \cite{Alexander:25} & Si & SFWM/MR & $\sim$2 & 0.995(3) & 0.2 & $-$ & $-$ \\[1.6pt]
Llewellyn \emph{et al.} \cite{Llewellyn:20} & Si & SFWM/MR & 5 & 0.72 & 0.5 & 0.683 & $-$ \\[1.6pt]
Chen \emph{et al.} \cite{ChenLZ:24} & Si & SFWM/MR & $-$ & 0.81 & 0.02 & 0.73 & 0.11 \\[1.6pt]
Bao \emph{et al.} \cite{Bao:23} & Si & SFWM/WG & Broadband & 0.78 & 0.02 & $-$ & $-$ \\[1.6pt]
Adcock \emph{et al.} \cite{Adcock:19} & Si & SFWM/WG & Broadband & 0.82 & 0.02 & 0.78 & 0.006 \\[1.6pt]
Lee \emph{et al.} \cite{Lee:24} & Si & SFWM/WG & Broadband & 0.96 & $-$ & 0.854 & 0.2 \\[1.6pt]
Paesani \emph{et al.} \cite{Paesani:20} & Si & SFWM/WG & Broadband & 0.96 & 0.006 & $-$ & $-$ \\[1.6pt]
Samara \emph{et al.} \cite{Samara:21} & Si$_3$N$_4$ & SFWM/MR & 0.3 & 0.932 & 0.13 & $-$ & $-$ \\[1.6pt]
Kuttner \emph{et al.} \cite{Kuttner:26} & TFLN & SPDC/WG & Broadband & 0.71 & $-$ & $-$ & $-$ \\[1.6pt]
Wang \emph{et al.} \cite{WangX:25} & TFLN & QD & $-$ & 0.73 & $-$ & $-$ & $-$ \\[1.6pt]
\end{tabular}
\end{ruledtabular}
\label{tab:S1}
\end{table*}

\noindent\textbf{Generation of four-photon GHZ state.}
We next demonstrate the generation of a four-photon GHZ state \cite{GHZ} by executing on-chip photonic qubit fusion between the two path-encoded EPR states, as schematically presented in Fig. \ref{Fig:4}a. 
Starting from the initial preparation of EPR states Q1--Q2 and Q3--Q4, the C29 photons enter the fusion circuit as qubits Q2 and Q3, where the logical $\ket{1}$ modes of Q2 and Q3 are spatially swapped. 
Successful post-selection, achieved by registering exactly one photon at each qubit output, projects qubits Q2 and Q3 onto the even-parity subspace via the operator $\ket{00}\bra{00}+\ket{11}\bra{11}$, yielding the four-photon GHZ state
\begin{equation}
    |G_4\rangle = \frac{1}{\sqrt{2}}\left(\ket{0}_1\ket{0}_2\ket{0}_3\ket{0}_4+\ket{1}_1\ket{1}_2\ket{1}_3\ket{1}_4\right),
\end{equation}
where the subscripts 1--4 index the output qubits Q1--Q4. 

In our experiment, each microresonator is driven at an on-chip pump power of approximately 0.2 mW, corresponding to an individual EPR state generation probability $p_\mathrm{EPR}=0.012$. 
We characterize the generated four-photon GHZ state by measuring its state fidelity $F$. 
For an $N$-photon GHZ state, the fidelity decomposes into two experimentally accessible components: 
\begin{equation}
F=\frac{1}{2}\langle P\rangle + \frac{1}{2}\langle C\rangle,
\end{equation}
where $P$ is the population and $C$ is the coherence. 
We first evaluate the population, defined as 
\begin{equation}
P=(\ket{0}\bra{0})^{\otimes N}+(\ket{1}\bra{1})^{\otimes N}.
\end{equation}
Figure \ref{Fig:4}b displays the fourfold counts accumulated over 10 s, yielding $\langle P\rangle=0.968(10)$.
This population is concentrated within the $\ket{0}^{\otimes4}$ and $\ket{1}^{\otimes4}$ configurations, with minimal parasitic contributions distributed across the remaining fourteen orthogonal computational-basis states.

The coherence $C$ quantifies the superposition between the $\ket{0}^{\otimes N}$ and $\ket{1}^{\otimes N}$ components through the off-diagonal elements of the system's density matrix.
To measure $C$, we project each of the four photons into the diagonal basis $\left(\ket{0}\pm e^{i\theta}\ket{1}\right)/\sqrt{2}$ and acquire all $2^4=16$ fourfold coincidence combinations.
As a representative example, Fig. \ref{Fig:4}c plots the fourfold counts within the $|\pm\rangle=\left(\ket{0}\pm\ket{1}\right)/\sqrt{2}$ basis, where multiphoton interference strongly concentrates the counts into even-parity outcomes. 
Using these collective coincidence counts, we calculate the expectation value $\langle M_\theta^{\otimes N}\rangle$, from which $\langle C\rangle$ is extracted via
\begin{equation}
C=\frac{1}{N}\sum_{k=0}^{N-1}(-1)^k M_{k\pi/N}^{\otimes N}.
\end{equation}
Figure \ref{Fig:4}d shows the measured $\langle M_{k\pi/4}^{\otimes 4}\rangle$ for $k=0,1,2,3$, yielding $\langle C\rangle=0.917(12)$.
Combining these values with $\langle P\rangle=0.968(10)$ results in a GHZ state fidelity of $F=0.943(8)$ at a mean fourfold count rate of 27 Hz. 
This record-high fidelity surpasses the genuine entanglement \cite{Guhne:09} threshold of $0.5$ by more than 55 standard deviations. 
The average fourfold count rate is calculated by summing the total fourfold counts recorded across the $P$ and $C$ characterizations and dividing by the total integration time (see Supplementary Information Note 2).
Attaining this level of fidelity requires precise, simultaneous spectral alignment of both the C29 and C41 resonances across all four high-$Q$ microresonators, highlighting the exceptional cross-wafer geometric uniformity of our Si$_3$N$_4$ fabrication process (see Methods and Supplementary Information Note 3).
 
Furthermore, we map the interference fringe of the four-photon GHZ state. 
Figure \ref{Fig:4}e tracks the measured four-photon expectation value $\langle M_{\theta}^{\otimes 4}\rangle$ against $\theta$ varied from $0$ to $\pi$.
The sinusoidal fit, proportional to $\cos 4\theta$, captures the fourfold increase in oscillation frequency that characterizes an $N=4$ GHZ state. 
This $4\theta$ phase dependence provides unambiguous evidence of four-photon coherence, underscoring the utility of our on-chip platform for deployment in entanglement-enhanced quantum metrology \cite{Giovannetti:11}.

\noindent\textbf{Conclusion and outlook.}
In summary, we have demonstrated a monolithic, ultralow-loss Si$_3$N$_4$ integrated photonic platform that unifies narrowband path-encoded EPR sources, qubit-fusion circuits, and reconfigurable state-analysis modules on a single chip. 
While individual benchmarks---such as a heralded HOM visibility of 0.990(6) and an EPR state fidelity of 0.9875(3)---match or exceed the top performance across alternative quantum photonic platforms \cite{Alexander:25}, the defining competitive edge of this architecture emerges at the system level. 
By leveraging this highly efficient, fully integrated functional stack, we synthesize a four-photon GHZ state with a record-high fidelity of 0.943(8). 
Concurrently, the platform delivers a fourfold count rate of 27 Hz, surpassing prior silicon-photonic implementations by more than two orders of magnitude, as summarized in the comprehensive platform comparison in Table \ref{tab:S1}. 
Under higher pump power, the chip yields a fourfold count rate of 134 Hz while preserving a GHZ state fidelity of 0.860(10), remaining well above the genuine entanglement threshold (see Supplementary Information Note 2). 
These achievements firmly establish foundry-grade ultralow-loss Si$_3$N$_4$ integrated photonics as a leading hardware architecture for scalable DV-QIP. 
  
Beyond the synthesis of multiphoton entanglement, our work provides a rigorous validation of a complete, monolithic, source--circuit--measurement architecture. 
By simultaneously optimizing narrowband photon-pair generation, near-unity photon indistinguishability, and ultralow propagation loss, this platform bypasses the rate--fidelity trade-off that typically constrains probabilistic linear optical systems. 
Maximizing the system-level collection efficiency preserves practical count rates at low pair-generation probabilities $p$, thereby suppressing multi-pair emission noise at its physical source.
Straightforward optimizations to off-chip collection and detection coupling could elevate these state generation rates into the kilohertz regime, directly rivaling state-of-the-art, bulk-optic, free-space configurations \cite{WangXL:16}.
A comprehensive loss budget detailing these pathways is provided in Supplementary Information Note 4.
The intersection of low-loss transmission and narrow photon linewidths makes this platform well suited for fiber-based qubit distribution and future coherent interfaces.
Future iterations reducing optical loss, optimizing dispersion, and enhancing coupling can push these microresonators deeper into the strongly over-coupled regime \cite{Pfeiffer:17b}, targeting loaded linewidths on the order of 10 MHz \cite{Chen:2024} to maximize both photon extraction efficiency and spectral matching with solid-state quantum memories.

Looking ahead, the geometric and structural uniformity inherent to this Si$_3$N$_4$ framework offers a clear path toward fully self-contained, field-deployable quantum systems. 
Heterogeneous or hybrid integration of chip-scale, low-noise semiconductor pump lasers \cite{Stern:18, Shen:20, Xiang:21, Sun:25} and waveguide-integrated superconducting nanowire single-photon detectors (SNSPDs) \cite{Pernice:12, Najafi:15} will eliminate volatile fiber-chip interfaces, drastically cutting system losses and unlocking the potential for multiplexed resource-state factories. 
By expanding the spatial multiplexing arrays and cascading successive on-chip fusion gates, this architecture can scale seamlessly from few-photon GHZ states to large-scale, highly entangled cluster and graph states. 
Our results thus establish a volume-manufacturable, wafer-scale photonic backbone capable of seamlessly generating, manipulating, routing, and measuring quantum states. 
Ultimately, this platform can bridge the gap between laboratory few-photon physics, distributed satellite-to-ground quantum networks, and fault-tolerant, measurement-based photonic quantum computers.

\setcounter{figure}{0} 
\renewcommand{\figurename}{Extended Data Figure}
\renewcommand{\thefigure}{\arabic{figure}}

\begin{figure*}[t!]
\centering
\includegraphics[width=\linewidth]{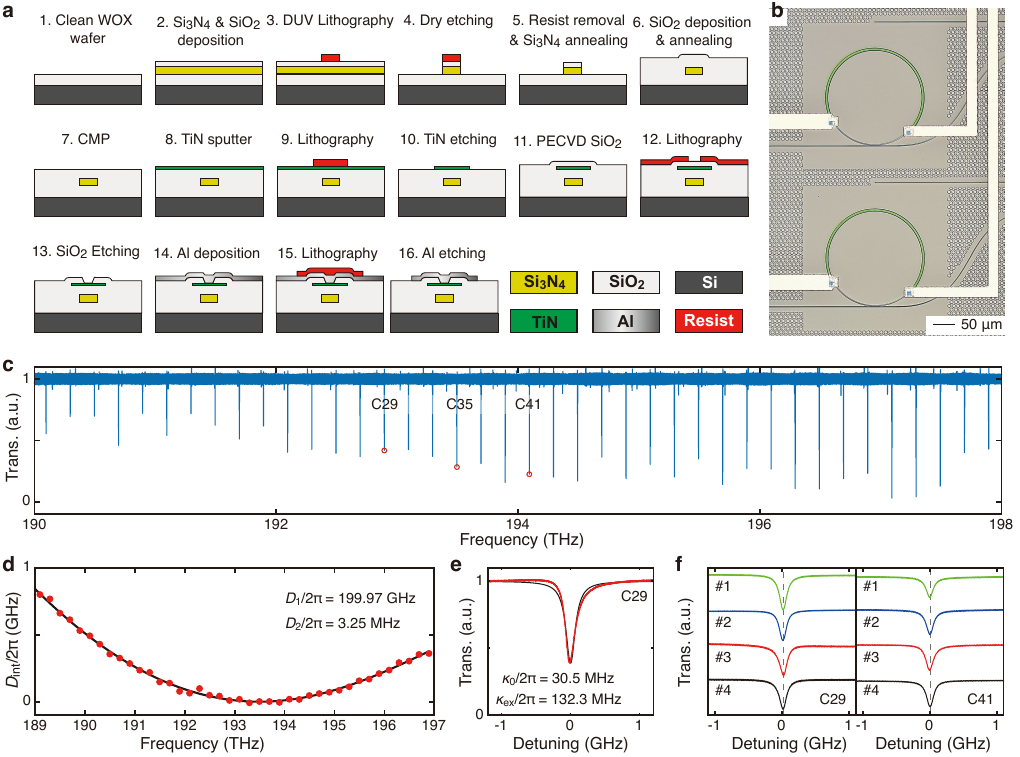}
\caption{
\textbf{Device fabrication process and characterization of the source microresonators.}
\textbf{a}, Schematic process flow for the Si$_3$N$_4$ PICs.
Fabrication starts with LPCVD deposition of Si$_3$N$_4$ and SiO$_2$ hardmask layers on a wet-thermal SiO$_2$/Si wafer, followed by DUV lithography, subtractive dry etching, high-temperature annealing, SiO$_2$ cladding deposition, and surface planarization.
Thermo-optic control is implemented by patterned TiN heaters embedded in PECVD SiO$_2$ and connected to Al metal routing through contact vias.
\textbf{b}, Optical micrograph of a fabricated device region with two microresonator-based photon-pair sources.
Green overlays indicate TiN heaters above the photonic waveguides, connected to the Al routing layer through etched contact vias.
\textbf{c}, Broadband transmission spectrum of a representative source microresonator over 190--198 THz.
\textbf{d}, Dispersion profile $D_\mathrm{int}/2\pi$ extracted from measured resonance frequencies.
The data are fitted with $D_\mathrm{int}=D_2\mu^2/2$, giving $D_1/2\pi=199.97$ GHz and $D_2/2\pi=3.25$ MHz.
\textbf{e}, Zoomed-in transmission spectrum of the C29 resonance with a Lorentzian fit.
The resonance is over-coupled, with $\kappa_0/2\pi=30.5$ MHz and $\kappa_\mathrm{ex}/2\pi=132.3$ MHz.
\textbf{f}, Simultaneous alignment of the C29 and C41 resonances across the four microresonators used to generate the four-photon GHZ state.
Fine thermal tuning aligns the resonances.
}
\label{ExdFig:1}
\end{figure*}

\medskip

\noindent \textbf{Methods}

\noindent\textbf{Device fabrication. }
The Si$_3$N$_4$ PICs are fabricated on 150-mm-diameter wafers in our CMOS foundry using an optimized deep-ultraviolet (DUV) subtractive process \cite{Liu:21, Ye:23}.
The process flow is illustrated in Extended Data Fig. \ref{ExdFig:1}a.
First, an 800-nm-thick Si$_3$N$_4$ film is deposited by low-pressure chemical vapor deposition (LPCVD) on a cleaned wet-thermal SiO$_2$/Si wafer.
An SiO$_2$ hardmask is then deposited on the Si$_3$N$_4$ layer, also by LPCVD.
DUV scanner lithography (ASML PAS 850C) is performed, followed by dry etching to transfer the pattern from the photoresist to the SiO$_2$ hardmask and then to the Si$_3$N$_4$ layer.
After photoresist removal, the wafer is thermally annealed at 1200 $^\circ$C in nitrogen.
A 3-$\mu$m-thick SiO$_2$ top cladding layer is deposited, followed by a second thermal anneal at 1200 $^\circ$C.
To fabricate the thermo-optic modulators, the wafer surface is planarized by chemical--mechanical polishing (CMP).
A titanium nitride (TiN) thin film is deposited and patterned to form resistive heaters above selected photonic structures.
The TiN heaters are embedded in an additional plasma-enhanced chemical vapor deposition (PECVD) SiO$_2$ layer.
Contact vias are opened through this oxide, and a top Al metallization layer connects the heaters to the wire-bond pads.
Finally, UV photolithography and deep dry etching create smooth chip facets for efficient fiber coupling.
The wafer is separated into individual chips by backside grinding or dicing.
Extended Data Fig. \ref{ExdFig:1}b shows an optical micrograph of a fabricated device region with two microresonator-based photon-pair sources.
The TiN heaters (shaded green) are positioned above the waveguides and electrically connected to the top Al routing layer through etched contact vias (shaded blue).

\noindent\textbf{Source microresonator characterization. }
The microresonators used for photon-pair generation have a radius of 114.5 $\mu$m, a Si$_3$N$_4$ thickness of 800 nm, and a waveguide width of 2 $\mu$m, corresponding to a free spectral range (FSR) of approximately 200 GHz.
Extended Data Fig. \ref{ExdFig:1}c shows a broadband transmission spectrum of the microresonators used in the experiment.
The extracted integrated-dispersion profile is shown in Extended Data Fig. \ref{ExdFig:1}d and defined as
\[
D_\mathrm{int}(\mu)\equiv \omega(\mu)-\omega_0-D_1\mu
= \frac{D_2}{2}\mu^2+\cdots,
\]
where $\omega(\mu)/2\pi$ is the resonance frequency of relative mode index $\mu$, with $\mu=0$ denoting the reference resonance at $\omega_0/2\pi$, $D_1/2\pi=199.97$ GHz is the FSR, and the positive $D_2/2\pi=3.25$ MHz corresponds to anomalous group velocity dispersion.
Higher-order dispersion terms are neglected in the fit.
Extended Data Fig. \ref{ExdFig:1}e shows a zoomed-in view of the C29 resonance, revealing a loaded linewidth of approximately 160 MHz.
In the broadband spectrum, the resonance dips deepen toward higher optical frequencies, indicating that C29 operates in the over-coupled regime.
Assuming unity coupling ideality \cite{Pfeiffer:17b}, the fit of the over-coupled branch gives $\kappa_0/2\pi=30.5$ MHz and $\kappa_\mathrm{ex}/2\pi=132.3$ MHz, corresponding to a photon extraction efficiency of approximately 80\%.

\noindent\textbf{Alignment of microresonators. } 
Each EPR source combines two microresonator photon-pair sources, and high-fidelity path-entanglement generation requires the corresponding photon pairs to be spectrally identical. 
In the GHZ state generation, this condition is met by simultaneously aligning the C29 and C41 resonances of all four high-$Q$ microresonators, as shown in Extended Data Fig.~\ref{ExdFig:1}f. 
This simultaneous alignment is enabled by the cross-wafer geometric uniformity of the fabrication process, with details provided in Supplementary Information Note 3. 
Fine tuning with integrated heaters provides the residual correction required for high-fidelity EPR state generation and high-quality photonic qubit fusion.

\noindent\textbf{Acknowledgments: }
We thank the engineering team of Qaleido Photonics for assisting this project. 
This project is supported by the National Key R\&D Program of China (Grant No. 2024YFA1409300), 
Quantum Science and Technology–National Science and Technology Major Project (Grant No. 2023ZD0301500), 
National Natural Science Foundation of China (Grant No. 12404417 and U25D9005), 
Shenzhen Science and Technology Program (Grant No. RCJC20231211090042078), 
and Shenzhen-Hong Kong Cooperation Zone for Technology and Innovation (HZQBKCZYB2020050). 
Y.-H. L. acknowledges support from the Young Elite Scientists Sponsorship Program by CAST (Grant No. YESS20240475). 

\noindent\textbf{Author contributions: }
Y.-H. L. and J. L. conceived the experiment. 
Y.-H. L. and R. C. designed the chips.
Z. Z., S. H., Z. S., Y. H., Z. C., S. L. and X. B. developed the fabrication process and fabricated the Si$_3$N$_4$ chip devices.
Y.-H. L., R. C., S. Z. and Y. C. built the setup and performed the experiment.
Y.-H. L., R. C. and J. L. analyzed the data and wrote the manuscript. 
J. L. supervised the project.

\noindent\textbf{Conflict of interest:} 
X. B. and J. L. are co-founders of Qaleido Photonics, a start-up that is developing silicon nitride integrated photonics technologies for foundry services. 
Others declare no conflicts of interest.

\noindent\textbf{Data availability: }
The code and data used to produce the plots within this work will be released in the repository Zenodo upon publication of this manuscript.

\bibliographystyle{apsrev4-1}
\bibliography{bibliography, bib_add}

\end{document}


\title{Supplementary Materials for: 
An ultralow-loss integrated photonic platform for discrete-variable quantum information processing}

\author{Ruiyang Chen}
\thanks{These authors contributed equally.}
\affiliation{International Quantum Academy and Shenzhen Futian SUSTech Institute for Quantum Technology and Engineering, Shenzhen 518048, China}
\affiliation{School of Physical Sciences and Hefei National Laboratory, University of Science and Technology of China, Hefei 230026, China}

\author{Zeying Zhong}
\thanks{These authors contributed equally.}
\affiliation{International Quantum Academy and Shenzhen Futian SUSTech Institute for Quantum Technology and Engineering, Shenzhen 518048, China}
\affiliation{Southern University of Science and Technology, Shenzhen 518055, China}

\author{Sanli Huang}
\affiliation{International Quantum Academy and Shenzhen Futian SUSTech Institute for Quantum Technology and Engineering, Shenzhen 518048, China}
\affiliation{School of Physical Sciences and Hefei National Laboratory, University of Science and Technology of China, Hefei 230026, China}

\author{Sicheng Zeng}
\affiliation{International Quantum Academy and Shenzhen Futian SUSTech Institute for Quantum Technology and Engineering, Shenzhen 518048, China}
\affiliation{Southern University of Science and Technology, Shenzhen 518055, China}

\author{Zhenyuan Shang}
\affiliation{International Quantum Academy and Shenzhen Futian SUSTech Institute for Quantum Technology and Engineering, Shenzhen 518048, China}
\affiliation{Southern University of Science and Technology, Shenzhen 518055, China}

\author{Yue Hu}
\affiliation{International Quantum Academy and Shenzhen Futian SUSTech Institute for Quantum Technology and Engineering, Shenzhen 518048, China}
\affiliation{Southern University of Science and Technology, Shenzhen 518055, China}

\author{Zhen Chen}
\affiliation{International Quantum Academy and Shenzhen Futian SUSTech Institute for Quantum Technology and Engineering, Shenzhen 518048, China}
\affiliation{School of Physical Sciences and Hefei National Laboratory, University of Science and Technology of China, Hefei 230026, China}

\author{Yuan Chen}
\affiliation{International Quantum Academy and Shenzhen Futian SUSTech Institute for Quantum Technology and Engineering, Shenzhen 518048, China}

\author{Shuyi Li}
\affiliation{International Quantum Academy and Shenzhen Futian SUSTech Institute for Quantum Technology and Engineering, Shenzhen 518048, China}

\author{Xue Bai}
\affiliation{International Quantum Academy and Shenzhen Futian SUSTech Institute for Quantum Technology and Engineering, Shenzhen 518048, China}
\affiliation{Qaleido Photonics, Shenzhen 518048, China}

\author{Yi-Han Luo}
\email{luoyh@iqasz.cn}
\affiliation{International Quantum Academy and Shenzhen Futian SUSTech Institute for Quantum Technology and Engineering, Shenzhen 518048, China}

\author{Junqiu Liu}
\email{liujq@iqasz.cn}
\affiliation{International Quantum Academy and Shenzhen Futian SUSTech Institute for Quantum Technology and Engineering, Shenzhen 518048, China}
\affiliation{School of Physical Sciences and Hefei National Laboratory, University of Science and Technology of China, Hefei 230026, China}

\maketitle
\tableofcontents
\clearpage


\section{Device details and experimental setup}
\vspace{0.2cm}

Our experiments use path-encoded photonic qubits, with $\ket{0}$ and $\ket{1}$ encoded in two distinct waveguide modes. 
Qubit manipulation is implemented through on-chip phase shift control, path exchange and coherent spatial mode mixing. 
These operations are based on building blocks including phase shifters, waveguide crossings and beam splitters (BSs). 
Together, they further form reconfigurable Mach--Zehnder interferometers (MZIs) and unbalanced Mach--Zehnder interferometers (UMZIs).
The chip also integrates photon-pair sources based on cavity-enhanced spontaneous four-wave mixing (SFWM) \cite{Helt:10} in high-$Q$ Si$_3$N$_4$ microresonators \cite{Chen:2024}. 
The microresonators and pump laser jointly determine the source linewidth, on-chip photon flux and indistinguishability. 
After on-chip processing and analysis, photons are coupled off-chip through edge couplers. 
They are filtered by dense wavelength-division multiplexing (DWDM) filters to suppress residual pump, before detection by superconducting nanowire single-photon detectors (SNSPDs).

Efficient and high-fidelity operation depends on the full experimental system, including the pump laser, photonic chip, chip--fiber coupling, filtering and detectors. 
This section describes the source microresonators, passive linear-optical components, thermo-optic tuning elements, chip--fiber coupling and experimental setup used for on-chip multiphoton interference and state characterization.

\subsection{Microresonators}

\begin{figure*}[b!]
\centering
\includegraphics[width=1.0\textwidth]{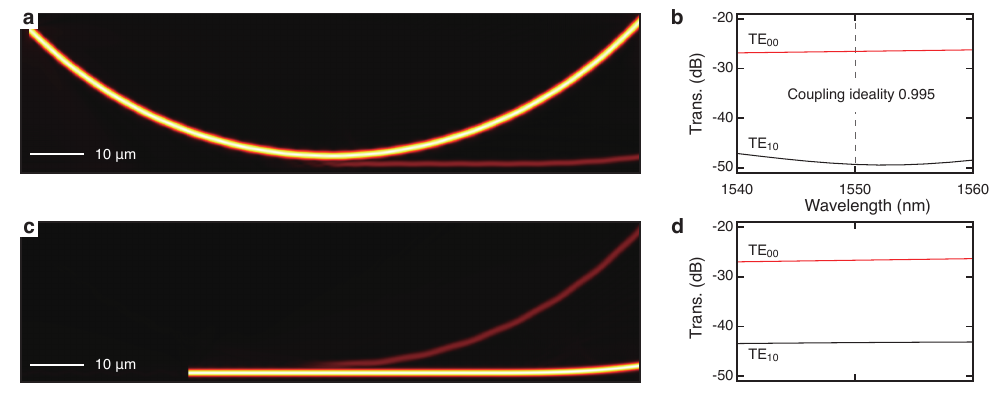}
\caption{
\textbf{Simulation of the source microresonator coupling region.}
\textbf{a} and \textbf{c}, Simulated optical field distribution for ring-to-bus out-coupling and bus-to-ring in-coupling at 1550~nm, respectively.
The field magnitude is plotted as $|E|^{1/2}$ to enhance visualization. 
\textbf{b}, Wavelength-dependent transmission into the TE$_{00}$ and TE$_{10}$ bus waveguide modes for the out-coupling.
The coupling ideality reaches 0.995 at 1550~nm.
\textbf{d}, Wavelength-dependent transmission into the TE$_{00}$ and TE$_{10}$ ring waveguide modes for the in-coupling.
The TE$_{10}$ mode is suppressed by more than 15~dB relative to the TE$_{00}$ mode.
}
\label{Fig:S-MR}
\end{figure*}

Photon pairs are generated by cavity-enhanced SFWM in Si$_3$N$_4$ microresonators.
The ring waveguides have an 800-nm-thick, 2-$\mu$m-wide Si$_3$N$_4$ core that provides anomalous group velocity dispersion for SFWM phase matching.
The microresonators have a radius of 114.5 $\mu$m, corresponding to a free spectral range (FSR) of approximately 200 GHz, which is compatible with the ITU frequency grid and allows standard fiber DWDM components to demultiplex the generated photons.

Each ring is side-coupled to a bus waveguide for pump injection and photon extraction.
The coupling region is designed to operate in the over-coupled regime while maintaining high coupling ideality \cite{Pfeiffer:17b}.
In this region, the ring waveguide is adiabatically tapered from 2 $\mu$m to 1.8 $\mu$m to increase the coupling strength to the bus waveguide.
The bus waveguide has a width of 1.4 $\mu$m, and the ring--bus gap is 500 nm.
This geometry provides mode-selective coupling for both ring-to-bus out-coupling and bus-to-ring in-coupling.

Figure \ref{Fig:S-MR} shows the simulated performance of the coupling region. 
The optical field distributions for out-coupling and in-coupling at 1550 nm are shown in Figs. \ref{Fig:S-MR}a and \ref{Fig:S-MR}c. 
The power transmissions from fundamental TE$_{00}$ mode to the TE$_{00}$ and TE$_{10}$ modes are shown in Figs. \ref{Fig:S-MR}b and \ref{Fig:S-MR}d. 
For out-coupling (Fig. \ref{Fig:S-MR}b), the simulated coupling ideality reaches 0.995 at 1550 nm. 
For in-coupling (Fig. \ref{Fig:S-MR}d), the power coupled to the TE$_{10}$ mode is more than 15 dB lower than that coupled to the TE$_{00}$ mode, confirming highly mode-selective excitation of the fundamental intracavity mode.

\begin{figure*}[b!]
\centering
\includegraphics[width=0.86\textwidth]{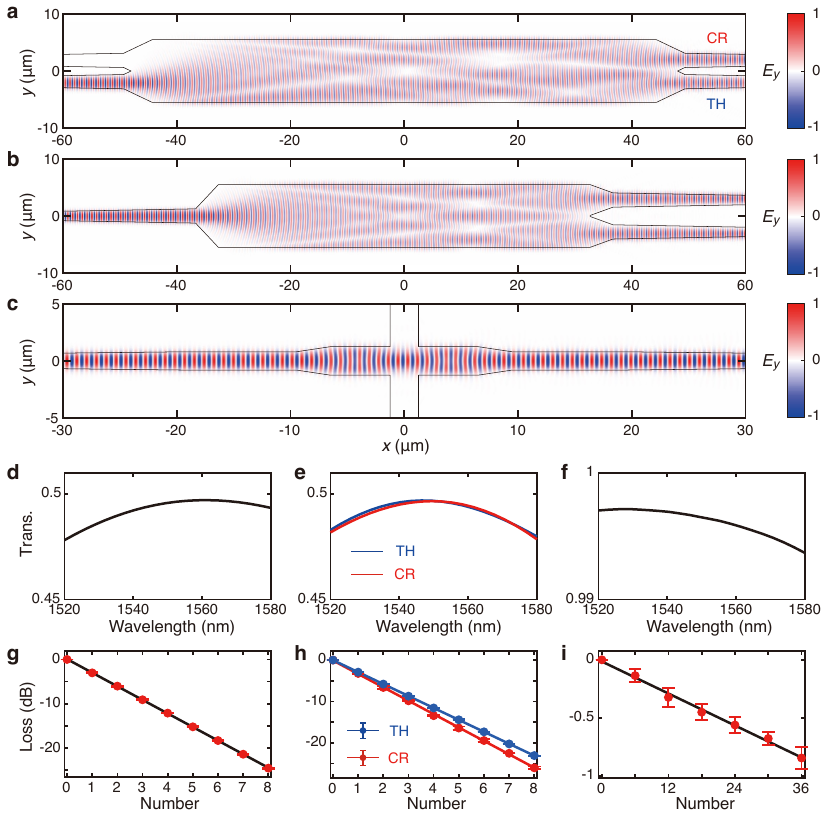}
\caption{
\textbf{Simulation and characterization of MMI-based components.}
\textbf{a}--\textbf{c}, Simulated $E_y$ field profile at 1550 nm for the $1\times2$ MMI, the $2\times2$ MMI and the waveguide crossing, respectively.
\textbf{d}--\textbf{f}, Simulated single-output transmission of the $1\times2$ MMI, the $2\times2$ MMI and the waveguide crossing over 1520--1580 nm, respectively.
\textbf{g}, Measured transmission of cascaded $1\times2$ MMIs, giving a single-output transmission of 0.494 and a summed two-output insertion loss of 0.053 dB per device.
\textbf{h}, Measured transmission of cascaded $2\times2$ MMIs, giving $T_\mathrm{CR}=0.474$ and $T_\mathrm{TH}=0.514$, corresponding to a summed two-output insertion loss of 0.052 dB per device.
\textbf{i}, Measured transmission of cascaded waveguide crossings.
A six-crossing unit has a transmission of 0.969, corresponding to a per-crossing insertion loss of 0.023 dB.
}
\label{Fig:S-MMI}
\end{figure*}

\subsection{Devices for photon splitting and routing}

Multiphoton manipulation on the chip uses three passive multimode interferometer (MMI) based building blocks: $1\times2$ MMIs, $2\times2$ MMIs and waveguide crossings.
These devices exploit self-imaging in a short multimode section \cite{Soldano:95}, providing broadband and fabrication-tolerant splitting, mode mixing and routing.
The $1\times2$ MMI is used as a balanced splitter, the $2\times2$ MMI serves as the BS for two-photon interference, and the crossing enables compact routing of path-encoded qubits.

Figures \ref{Fig:S-MMI}a--c show the simulated $E_y$ field profiles at 1550 nm for the $1\times2$ MMI, $2\times2$ MMI and waveguide crossing.
The corresponding simulated spectral responses over 1520--1580 nm are shown in Figs. \ref{Fig:S-MMI}d--f.
Both MMI splitters maintain nearly wavelength-independent balanced operation across the telecom C band, while the crossing has a simulated transmission above 0.99 throughout the same wavelength range.

The fabricated devices are characterized using cascaded test structures across multiple chips.
The per-device transmission is extracted from a linear fit to the measured transmission versus the number of cascaded elements, as shown in Fig. \ref{Fig:S-MMI}g--i.
For the $1\times2$ MMI, the measured single-output transmission is 0.494, corresponding to a summed two-output transmission of 0.988 and an insertion loss of 0.053~dB.
For the $2\times2$ MMI, the measured cross- and through-port transmissions are $T_\mathrm{CR}=0.474$ and $T_\mathrm{TH}=0.514$, respectively.
The summed transmission is therefore 0.988, giving an insertion loss of 0.052~dB and an effective splitting ratio of $48.0/52.0$ after normalization by the total transmission.
For the waveguide crossing, a 6-crossing unit has a measured transmission of 0.969, indicating a per-crossing transmission of 0.995 and a per-crossing insertion loss of 0.023~dB.

\subsection{Interferometers and their thermal tuning}
\begin{figure*}[b!]
\centering
\includegraphics[width=0.88\textwidth]{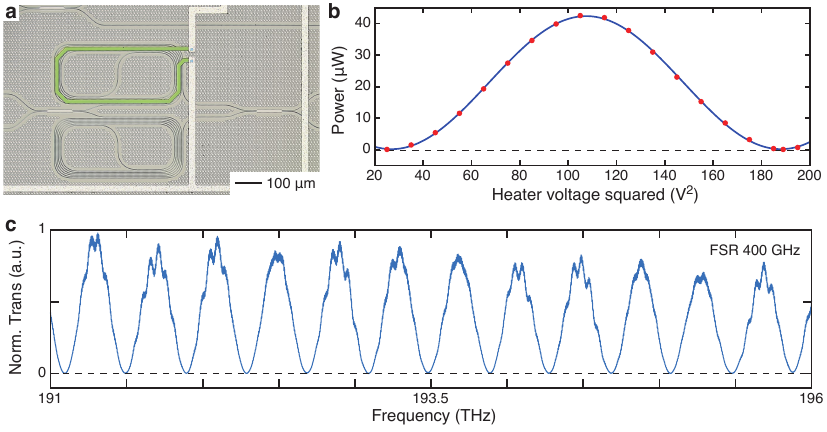}
\caption{
\textbf{Performance of Mach--Zehnder interferometers and unbalanced Mach--Zehnder interferometers.}
\textbf{a}, Optical microscope image of a representative MZI.
\textbf{b}, Measured output power as a function of squared heater voltage.
The red points are experimental data and the blue curve is a sinusoidal fit, showing smooth thermal tuning of the interferometer phase.
\textbf{c}, Broadband transmission spectrum of a representative UMZI.
The measured spectral period corresponds to an FSR of approximately $400$ GHz.
}
\label{Fig:S-UMZI}
\end{figure*}

MZIs serve as reconfigurable interference elements in the photonic circuit, providing frequency demultiplexing, path routing and qubit analysis.
In our experiments, the relative phase between the two arms is tuned by an integrated resistive heater on one arm, as shown in Fig. \ref{Fig:S-UMZI}a.
The phase shift is approximately proportional to the electrical power, thus the tuning response is characterized as a function of the squared heater voltage.
Figure \ref{Fig:S-UMZI}b shows a representative thermal tuning curve.
The measured optical power follows a sinusoidal dependence on the squared heater voltage, confirming coherent phase control over a range exceeding $2\pi$.
The smooth interference fringe indicates stable operation of the thermal phase shifter. 

For frequency demultiplexing, we use UMZIs with a designed FSR matched to the frequency-bin spacing required by the source and filtering architecture.
Figure \ref{Fig:S-UMZI}c shows the broadband transmission spectrum of a representative UMZI.
A periodic spectral response with an FSR of approximately $400$ GHz is observed across the telecom C band.

\subsection{Chip--fiber edge coupling}

Both pump injection and photon collection are implemented through edge coupling at the chip facets.
Multi-channel input and output coupling to the photonic integrated circuit (PIC) is provided by V-groove-mounted UHNA-4 fiber arrays with a 127 $\mu$m pitch. 
The UHNA-4 fibers are thermally expanded and spliced to standard SMF-28e fibers for connection to the subsequent fiber components.
On-chip inverse tapers with a 240-nm tip width are used as edge couplers, matching the UHNA-4 fiber mode profile.
The fiber arrays are aligned to the on-chip tapers with sub-micrometer precision using DC-motor-driven Hexapod stages (Physik Instrumente H-811.I2).
With an index-matching liquid of refractive index 1.50 applied at the chip--fiber interface, the average chip--fiber coupling efficiency is approximately 70\%, corresponding to a coupling loss of 1.55 dB/facet.

\subsection{Experimental setup}

\begin{figure*}[b!]
\centering
\includegraphics[width=1.0\textwidth]{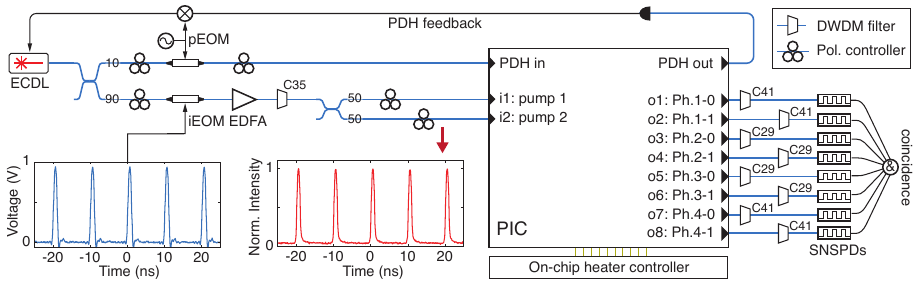}
\caption{
\textbf{Experimental setup.}
Schematic of the optical, electrical and detection setup.
A fraction of the external-cavity diode laser is used for PDH locking to a reference microresonator on the PIC, while the remaining light is carved into pump pulses by an iEOM, amplified by an EDFA, spectrally filtered and split into two pump inputs.
The generated photons are demultiplexed by DWDM filters and detected by SNSPDs.
All on-chip microresonators and interferometers are controlled by integrated heaters.
}
\label{Fig:S-setup}
\end{figure*}

The experimental setup is shown in Fig.~\ref{Fig:S-setup}.
An external-cavity diode laser (ECDL, Toptica CTL 1550) provides the pump light.
The laser output is first divided into two branches.
The 10\% branch is used for frequency stabilization.
This light is phase-modulated by a phase electro-optic modulator (pEOM) to generate the sidebands required for Pound--Drever--Hall (PDH) locking, and is then sent to the \emph{PDH in} port of the PIC.
The \emph{PDH in} port is connected to a reference microresonator fabricated on the same chip and identical to the photon-pair-generation microresonators.
The transmitted light from the \emph{PDH out} port is detected and processed to generate the PDH error signal.
This error signal is fed back to the ECDL to lock the laser frequency to the reference microresonator resonance.

The remaining 90\% branch generates the pulsed pump for photon-pair generation.
The continuous-wave laser is carved into pulses by an intensity electro-optic modulator (iEOM).
The iEOM is biased at the minimum-transmission point and driven by a periodic electrical pulse train, as shown by the blue trace in the inset of Fig. \ref{Fig:S-setup}.
After the iEOM, the optical pulses are amplified by an erbium-doped fiber amplifier (EDFA), producing the pulse train shown by the red trace in the inset.
The pulse repetition rate is 100 MHz and the pulse duration is approximately 0.9 ns.
A DWDM filter centered at the C35 channel, corresponding to 193.5 THz, is inserted after the EDFA to suppress out-of-band amplified spontaneous emission by approximately 100 dB.

The filtered pump pulses are equally split and injected into the \emph{i1: pump 1} and \emph{i2: pump 2} ports of the PIC.
These two pump inputs drive two independent on-chip EPR sources.
Photons are then routed, interfered and analyzed by the on-chip linear optical circuit.
The output photons are collected from the eight output ports \emph{o1}--\emph{o8}, corresponding to the two path modes of four path-encoded photonic qubits.
Before detection, each output channel is spectrally filtered by a DWDM filter matched to either the C29 or C41 channel, which removes residual pump light and selects the desired signal or idler photons.
The photons are finally detected by SNSPDs, and all detection events are recorded by a time tagger (Swabian Time Tagger Ultra) for coincidence analysis.

The five microresonators, comprising one reference microresonator for PDH locking and four microresonators for photon-pair generation, are thermally tuned by integrated heaters.
The MZIs and UMZIs are controlled in the same way.
Each heater is driven by an independent voltage source with a 0--20 V tuning range and 16-bit voltage resolution.
The microresonator heaters are used to align the relevant resonances involved in entangled states generation and HOM interference.
The heaters on the 800-GHz-FSR UMZIs are used to demultiplex the C29 and C41 frequency channels, while the MZI heaters are used to configure the measurement bases for state analysis.
The chip temperature is stabilized to within approximately 1 mK.
Upon the laser is locked to the reference microresonator, all four high-$Q$ microresonators are aligned to the pump frequency by their on-chip heaters.
No additional active feedback is applied to the source microresonators during the experiments.

\section{Manipulation of path-encoded photonic qubits}
\vspace{0.2cm}

\subsection{Hong--Ou--Mandel interference and photon indistinguishability}

Near-unity photon indistinguishability is required for high-fidelity photonic qubit fusion. 
For a parametric photon-pair source, the indistinguishability of heralded photons is often discussed in terms of spectral purity, which can be inferred from the non-heralded second-order correlation $g_{\rm nh}^{(2)}(0)$ \cite{Christ:11}.
This metric has two limitations for integrated photonic platforms.
First, unavoidable nonlinear scattering, such as Raman scattering, introduces additional thermal components into the non-heralded photon statistics, thereby reducing the spectral purity inferred from $g_{\rm nh}^{(2)}(0)$. 
However, because these noise photons are largely uncorrelated with the heralding photons, they are strongly rejected by the heralding condition and do \emph{not} contribute appreciably to the post-selected multiphoton interference. 
Second, $g_{\rm nh}^{(2)}(0)$ characterizes only the modal structure of one single photon-pair source.
It does not capture source-to-source mismatch, such as differences in resonance frequency, linewidth, and spectral distribution between independent sources.
Such mismatch directly reduces multiphoton interference visibility even when each individual source has high spectral purity.
We therefore directly benchmark the experimentally relevant photon indistinguishability by measuring the HOM interference visibility between heralded single photons from two independent sources.

In this section, we first analyze how spectral-mode overlap and spectral purity jointly determine the HOM interference visibility. 
We then correct for the visibility reduction caused by the imbalance of the BS, which must be de-embedded to extract the intrinsic source indistinguishability. 
Finally, we present the experimental procedure to measure HOM interference visibility in our experiment.

We first consider the case where two pure single photons are incident on the two input ports, labelled 1 and 2, of a lossless 50:50 BS.
Generally, the two photons may have different spectral distributions.
Thus the input state is written as
\begin{equation}
    \ket{\psi_{\rm in}}=\hat A_1^\dagger(f)\hat A_2^\dagger(g)\ket{\mathrm{vac}},
\end{equation}
where
\begin{equation}
\hat A_1^\dagger(f)=\int d\omega\, f(\omega)\hat a_1^\dagger(\omega), ~~\hat A_2^\dagger(g)=\int d\omega\, g(\omega)\hat a_2^\dagger(\omega), 
\label{eqn:photon-spec}
\end{equation}
with $\hat a_{1,2}^\dagger$ denoting the creation operators for the photons in input ports 1 and 2. 
Here $f(\omega)$ and $g(\omega)$ are the normalized spectral amplitudes of the two photons, with $\int d\omega |f(\omega)|^2=\int d\omega |g(\omega)|^2=1$.
Using the BS transformation \cite{Zeilinger:81, Pan:12}
\[
\hat a_1^\dagger(\omega)\rightarrow\frac{i\hat b_1^\dagger(\omega)+\hat b_2^\dagger(\omega)}{\sqrt{2}},
\qquad
\hat a_2^\dagger(\omega)\rightarrow\frac{\hat b_1^\dagger(\omega)+i\hat b_2^\dagger(\omega)}{\sqrt{2}},
\]
with $\hat b_{1,2}^\dagger$ denoting the creation operators for the photons in output ports 1 and 2. 
The reflection of BS preserves the spatial-port label and contributes a phase $\pi/2$, while its transmission switches the port label.
Thus the coincidence component from the output is calculated as
\begin{equation}
\ket{\psi_{\rm coin}}=\frac{1}{2}\int d\omega_1 d\omega_2\,
\left[f(\omega_2)g(\omega_1)-f(\omega_1)g(\omega_2)\right]
\hat b_1^\dagger(\omega_1)\hat b_2^\dagger(\omega_2)
\ket{\mathrm{vac}},
\end{equation}
with an amplitude antisymmetric under exchange of the two port labels.

If the two incident photons are spectrally identical, $f(\omega)\equiv g(\omega)$, the coincidence component vanishes exactly as $\ket{\psi_{\rm coin}}=0$.
This is the Hong--Ou--Mandel effect: two indistinguishable photons bunch into the same output port of the BS and eliminate the coincidence events.
For non-identical photons, the coincidence probability is obtained from the norm of $\ket{\psi_{\rm coin}}$ as
\[
P_{\rm c}=\langle\psi_{\rm coin}|\psi_{\rm coin}\rangle
=\frac{1}{4}\int d\omega_1 d\omega_2\,\left|f(\omega_1)g(\omega_2)-f(\omega_2)g(\omega_1)\right|^2.
\]
Using the normalization of $f$ and $g$, this becomes
\[
P_{\rm c}=\frac{1}{2}\left(1-\left|\int d\omega\, f^*(\omega)g(\omega)\right|^2\right).
\]
The ideal HOM interference visibility is therefore
\begin{equation}
V_{\rm HOM}=\left|\int d\omega\, f^*(\omega)g(\omega)\right|^2.
\end{equation}

We next consider heralded photons from parametric photon-pair sources, where each photon typically occupies a mixed spectral state, described by an incoherent mixture of mutually orthogonal spectral modes
\[
\hat\rho_1=\sum_m p_m\ket{u_m}\bra{u_m},\quad
\hat\rho_2=\sum_n q_n\ket{v_n}\bra{v_n},
\]
where $\ket{u_m}$ and $\ket{v_n}$ denote normalized spectral modes, while $p_m$ and $q_n$ are their classical probabilities, satisfying $\braket{u_m}{u_{m'}}=\delta_{mm'}$, $\braket{v_n}{v_{n'}}=\delta_{nn'}$ and $\sum_m p_m=\sum_n q_n=1$. 
For a given mode pair $\ket{u_m}$ and $\ket{v_n}$, the HOM interference visibility is $V_{mn}=|\braket{u_m}{v_n}|^2$. 
Thus for the statistical mixtures, the observed visibility is the probability-weighted average over all mode-pair visibilities,
\begin{equation}
V_\mathrm{HOM} = \sum_{m,n} p_m q_n |\braket{u_m}{v_n}|^2, 
\label{eqn:HOM-vis}
\end{equation}
which can be further written in density-matrix form as $V_\mathrm{HOM}=\mathrm{Tr}\left(\hat\rho_1\hat\rho_2\right)$.
Equation \eqref{eqn:HOM-vis} shows that the HOM interference visibility is affected by both state purity and spectral-mode overlap.
If both photons are pure but occupy different modes, the visibility is reduced to $V_{\rm HOM}=|\braket{u}{v}|^2$.
If the two photons have identical mixed spectral states, namely $\hat\rho_1=\hat\rho_2=\hat\rho$, the visibility is limited by the purity of the single-photon state $V_{\rm HOM}=\mathrm{Tr}\left(\hat\rho^2\right)$.
Specifically, for $\hat\rho=\sum_m p_m\ket{u_m}\bra{u_m}$, we have $V_{\rm HOM}=\sum_m p_m^2$. 

\begin{figure*}[t!]
\centering
\includegraphics[width=0.85\textwidth]{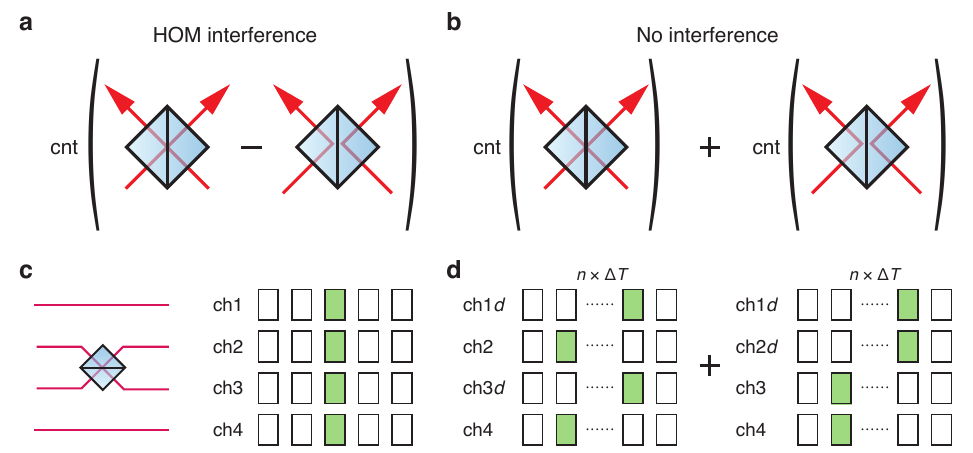}
\caption{
\textbf{Hong--Ou--Mandel interference visibility measurement.}
\textbf{a}, For indistinguishable heralded photons incident on the BS, the two coincidence amplitudes corresponding to both photons being transmitted and both photons being reflected interfere destructively, suppressing the fourfold coincidence events.
\textbf{b}, For distinguishable photons, the two alternatives do not interfere and their probabilities add incoherently, giving the non-interfering coincidence event.
\textbf{c}, Experimental channel assignment for the on-chip HOM measurement.
Channels 1 and 4 detect the idler photons and serve as heralding channels, while channels 2 and 3 detect the two signal photons after the BS.
Green boxes indicate the same pump-pulse time bin used to register the interfering fourfold event $c(1,2,3,4)$.
\textbf{d}, Virtual delayed-channel method used to obtain the non-interfering baseline without physically scanning the optical delay.
The time tagger generates delayed channels $1d$, $2d$, and $3d$ by shifting the corresponding time tags by $n\Delta T$, where $\Delta T=10$ ns is the laser pulse period.
The two delayed fourfold combinations, $c(1d,2,3d,4)$ and $c(1d,2d,3,4)$, measure the two non-interfering contributions in parallel with the interfering fourfold measurement.
}
\label{Fig:S-HOM}
\end{figure*}

In practice, the measured HOM interference visibility is also affected by beam-splitter imbalance.
For a BS with reflection $R$ and transmission $T$, its transformation is
\[
\hat a_1^\dagger(\omega)\rightarrow\sqrt{T}\hat b_2^\dagger(\omega)+i\sqrt{R}\hat b_1^\dagger(\omega),\text{ and }\hat a_2^\dagger(\omega)\rightarrow i\sqrt{R}\hat b_2^\dagger(\omega)+\sqrt{T}\hat b_1^\dagger(\omega).
\]
Thus for general mixed input states, the coincidence probability is
\[
P_{\rm c}=R^2+T^2-2RT\,V_\mathrm{HOM},
\]
where $V_\mathrm{HOM}$ is defined as Eq. \eqref{eqn:HOM-vis}. 
When the two photons are fully distinguishable, which can be experimentally realized by introducing a relative temporal delay, we have $V_\mathrm{HOM}=0$.
The coincidence probability then reaches the distinguishable-photon baseline $P_{\rm c}^{\rm dist}=R^2+T^2$. 
Using the measured values $P_{\rm c}$ and $P_{\rm c}^{\rm dist}$, $V_\mathrm{HOM}$ is extracted as
\begin{equation}
V_{\rm HOM} = \left(1-\frac{P_{\rm c}}{P_{\rm c}^{\rm dist}}\right) \left/ \frac{2RT}{R^2+T^2} \right.,
\end{equation}
which gives the photon indistinguishability after de-embedding the measured beam-splitter imbalance.

In our experiment, HOM interference is implemented on chip, so $P_{\rm c}^{\rm dist}$ cannot be obtained by inserting a meter-scale relative path delay. 
Instead, we use the time tagger's virtual-delay channels to record interfering and non-interfering fourfold coincidence events in parallel, corresponding to the conditions shown in Figs.~\ref{Fig:S-HOM}a and \ref{Fig:S-HOM}b, respectively.
Experimentally, as shown in Fig.~\ref{Fig:S-HOM}c, the photons in channel C41 from the two independent photon-pair sources are detected in channels 1 and 4 and serve as heralding triggers.
The two heralded C29 photons interfere on the BS and are detected in channels 2 and 3.
In addition, the time tagger generates virtual delayed channels $1d$, $2d$, and $3d$ by applying an internal delay of $n\times\Delta T$ to channels 1, 2, and 3, respectively, where $\Delta T=10$ ns is the time interval between two adjacent laser pulses, and $n=0, \pm1, \pm2, \cdots$ is an integer.  

The fourfold coincidence events among channels 1, 2, 3 and 4 correspond to simultaneous arrival and interference of the two C29 photons, giving the interfering fourfold count, denoted by $c(1,2,3,4)$.
By introducing the virtual delay, the coincidences are formed from different pump pulses, and HOM interference is removed.
As shown in Fig.~\ref{Fig:S-HOM}d, the coincidence events among channels $1d$, 2, $3d$ and 4 correspond to both C29 photons being transmitted, while the coincidence events among channels $1d$, $2d$, 3 and 4 correspond to both C29 photons being reflected.
These two non-interfering contributions are denoted by $c(1d,2,3d,4)$ and $c(1d,2d,3,4)$, respectively.
The HOM interference visibility is then extracted with
\begin{equation}
V_\mathrm{HOM} = \left(1 - \frac{c(1,2,3,4)}{c(1d,2,3d,4) + c(1d,2d,3,4)}\right)\left/\frac{2RT}{R^2+T^2} \right..
\end{equation}

\subsection{Path-encoded EPR state generation}

The main text describes the path-exchange scheme used to generate a path-encoded EPR state. 
To quantify how resonance misalignment affects the EPR state coherence, we retain the spectral degrees of freedom.
Denoting the joint spectral amplitudes (JSAs) of the two photon-pair sources by $F(\omega_1,\omega_2)$ and $G(\omega_1,\omega_2)$, the generated EPR state can be written as
\begin{equation}
\ket{\Psi} = \frac{1}{\sqrt{N}}\int d\omega_1 d\omega_2
\left[F(\omega_1,\omega_2)\hat{a}_0^\dagger(\omega_1)\hat{b}_0^\dagger(\omega_2)
+G(\omega_1,\omega_2)\hat{a}_1^\dagger(\omega_1)\hat{b}_1^\dagger(\omega_2)
\right]\ket{\mathrm{vac}},
\end{equation}
where $\hat{a}^\dagger$ and $\hat{b}^\dagger$ are the creation operators for the two photons, and the subscripts $0$ and $1$ denote the two path-encoded logical modes. 
The normalization factor is $N=A+B$, where
\[
A=\int d\omega_1 d\omega_2 |F(\omega_1,\omega_2)|^2,
\quad
B=\int d\omega_1 d\omega_2 |G(\omega_1,\omega_2)|^2.
\]

After tracing over the spectral degrees of freedom, the reduced density matrix of the path state, written in the basis $\{\ket{00}, \ket{01}, \ket{10},\ket{11}\}$, is
\begin{equation}
\rho_{\mathrm{path}} = \frac{1}{A+B}
\begin{pmatrix}
A & 0 & 0 & \Lambda \\
0 & 0 & 0 & 0 \\
0 & 0 & 0 & 0 \\
\Lambda^* & 0 & 0 & B
\end{pmatrix}, 
\label{eqn:EPR-den}
\end{equation}
with
\[
\Lambda=\int d\omega_1 d\omega_2 F(\omega_1,\omega_2)G^*(\omega_1,\omega_2).
\]
The parameters $A$ and $B$ describe the population balance components $\ket{00}$ and $\ket{11}$, while $\Lambda$ determines their superposition coherence. 

To connect this density-matrix description with the experimental measurements, we consider $\langle X\otimes X\rangle$ as an example. 
Substituting Eq. \eqref{eqn:EPR-den} gives
\[
\langle X\otimes X\rangle=\frac{2\mathrm{Re}(\Lambda)}{A+B}.
\]
For the target state $(\ket{00}+\ket{11})/\sqrt{2}$, the relative phase between $\ket{00}$ and $\ket{11}$ is set to zero, so $\Lambda=\mathrm{Re}(\Lambda)=|\Lambda|$.
Defining the normalized JSA overlap as $\mu=|\Lambda|/\sqrt{AB}$, we obtain
\begin{equation}
\langle X\otimes X\rangle=\frac{2\sqrt{AB}}{A+B}\mu.
\label{eqn:EPR-XX}
\end{equation}
Equation \eqref{eqn:EPR-XX} shows that $\langle X\otimes X\rangle$ is jointly determined by the population balance and the JSA overlap of the two photon-pair sources. 
It reaches unity only for balanced populations, $A=B$, and identical photon-pair JSAs, $\mu=1$.

\subsection{Fusing two EPR states to form a four-photon GHZ state}

We next consider GHZ state generation by fusing two path-encoded EPR states. 
The experiment involves four microresonator photon-pair sources with JSAs $F_1$, $G_1$, $F_2$ and $G_2$. 
Here, $F_1$ and $G_1$ describe the two path components of the first EPR state, while $F_2$ and $G_2$ describe those of the second EPR state. 
The first EPR state is carried by photons in modes $a$ and $b$, and the second by photons in modes $c$ and $d$. The initial unnormalized two-EPR state can then be written as
\begin{equation}
\ket{\widetilde \Psi_{2{\rm EPR}}}=
\left(\hat S_{F_1}^{\dagger}+\hat S_{G_1}^{\dagger}\right)
\left(\hat S_{F_2}^{\dagger}+\hat S_{G_2}^{\dagger}\right)
\ket{\mathrm{vac}}, 
\label{eqn:GHZ-init1}
\end{equation}
where
\begin{subequations}
\begin{align}
\hat S_{F_1}^{\dagger}&=\int d\omega_a d\omega_b\,F_1(\omega_a,\omega_b)\hat a_0^\dagger(\omega_a)\hat b_0^\dagger(\omega_b),
\\
\hat S_{G_1}^{\dagger}&=\int d\omega_a d\omega_b\,G_1(\omega_a,\omega_b)\hat a_1^\dagger(\omega_a)\hat b_1^\dagger(\omega_b),\\
\hat S_{F_2}^{\dagger}&=\int d\omega_c d\omega_d\,F_2(\omega_c,\omega_d)\hat c_0^\dagger(\omega_c)\hat d_0^\dagger(\omega_d),\\
\hat S_{G_2}^{\dagger}&=\int d\omega_c d\omega_d\,G_2(\omega_c,\omega_d)\hat c_1^\dagger(\omega_c)\hat d_1^\dagger(\omega_d),
\end{align}
\end{subequations}
with subscripts $0$ and $1$ denoting the two spatial modes used for path encoding.
Expanding \eqref{eqn:GHZ-init1} gives
\begin{align}
\ket{\Psi_{2{\rm EPR}}}
=&
\frac{1}{\sqrt{N_1N_2}}\int d\omega_a d\omega_b d\omega_c d\omega_d\,
\Big[
F_1(\omega_a,\omega_b)F_2(\omega_c,\omega_d)
\hat a_0^\dagger(\omega_a)
\hat b_0^\dagger(\omega_b)
\hat c_0^\dagger(\omega_c)
\hat d_0^\dagger(\omega_d)
\nonumber\\
&+
F_1(\omega_a,\omega_b)G_2(\omega_c,\omega_d)
\hat a_0^\dagger(\omega_a)
\hat b_0^\dagger(\omega_b)
\hat c_1^\dagger(\omega_c)
\hat d_1^\dagger(\omega_d)
\nonumber\\
&+
G_1(\omega_a,\omega_b)F_2(\omega_c,\omega_d)
\hat a_1^\dagger(\omega_a)
\hat b_1^\dagger(\omega_b)
\hat c_0^\dagger(\omega_c)
\hat d_0^\dagger(\omega_d)
\nonumber\\
&+
G_1(\omega_a,\omega_b)G_2(\omega_c,\omega_d)
\hat a_1^\dagger(\omega_a)
\hat b_1^\dagger(\omega_b)
\hat c_1^\dagger(\omega_c)
\hat d_1^\dagger(\omega_d)
\Big]
\ket{\mathrm{vac}},
\label{eqn:two-epr-before-fusion}
\end{align}
where $N_1=A_1+B_1$ and $N_2=A_2+B_2$ are the normalization factors of the two individual EPR states, with
\begin{align*}
A_1 &= \int d\omega_a d\omega_b\, |F_1(\omega_a,\omega_b)|^2,
\quad B_1 = \int d\omega_a d\omega_b\, |G_1(\omega_a,\omega_b)|^2,
\\
A_2 &= \int d\omega_c d\omega_d\, |F_2(\omega_c,\omega_d)|^2,
\quad B_2 = \int d\omega_c d\omega_d\, |G_2(\omega_c,\omega_d)|^2.
\end{align*}

Photons in modes $b$ and $c$ are then fused. 
The fusion operation leaves the logical-$0$ modes unchanged and exchanges the logical-$1$ modes, with the corresponding mode transformation given by
\begin{align*}
\hat b_0^\dagger(\omega) \rightarrow \hat b_0^\dagger(\omega),\,
\hat c_0^\dagger(\omega) \rightarrow \hat c_0^\dagger(\omega),\,
\hat b_1^\dagger(\omega) \rightarrow \hat c_1^\dagger(\omega),\,
\hat c_1^\dagger(\omega) \rightarrow \hat b_1^\dagger(\omega).
\end{align*}
Successful fusion is post-selected by requiring exactly one photon in each fused output mode, $b$ and $c$. 
Writing $d\Omega=d\omega_a d\omega_b d\omega_c d\omega_d$, the post-selected state is
\begin{align}
\ket{\widetilde \Psi_{\rm GHZ}}=\frac{1}{\sqrt{N_1N_2}}\int d\Omega\,
\Big[
&
\mathcal F(\omega_a,\omega_b,\omega_c,\omega_d)
\hat a_0^\dagger(\omega_a)
\hat b_0^\dagger(\omega_b)
\hat c_0^\dagger(\omega_c)
\hat d_0^\dagger(\omega_d)
\nonumber\\
+&
\mathcal G(\omega_a,\omega_b,\omega_c,\omega_d)
\hat a_1^\dagger(\omega_a)
\hat b_1^\dagger(\omega_b)
\hat c_1^\dagger(\omega_c)
\hat d_1^\dagger(\omega_d)
\Big]
\ket{\mathrm{vac}},
\label{eqn:post-selected-ghz-state}
\end{align}
where the four-photon spectral amplitudes are
\begin{subequations}
\begin{align}
\mathcal F(\omega_a,\omega_b,\omega_c,\omega_d)
&=F_1(\omega_a,\omega_b)F_2(\omega_c,\omega_d),
\label{eqn:calF}
\\
\mathcal G(\omega_a,\omega_b,\omega_c,\omega_d)
&=G_1(\omega_a,\omega_c)G_2(\omega_b,\omega_d).
\label{eqn:calG}
\end{align}
\end{subequations}
The corresponding fusion success probability is
\begin{equation}
    p_{\rm fuse} = \braket{\widetilde \Psi_{\rm GHZ}}{\widetilde \Psi_{\rm GHZ}} = \frac{P_0+P_1}{N_1N_2}
\end{equation}
where $P_0 = A_1A_2$ and $P_1 = B_1B_2$.
For balanced EPR components, $A_1=B_1$ and $A_2=B_2$, the fusion success probability reduces to $p_{\rm fuse}=1/2$. 
Therefore, the normalized post-selected state is
\[
\ket{\Psi_{\rm GHZ}}=\sqrt{\frac{N_1N_2}{P_0+P_1}}\ket{\widetilde \Psi_{\rm GHZ}}.
\]

Tracing over all spectral degrees of freedom gives the reduced path state density operator as
\begin{align}
\rho_{\rm path}^{\rm GHZ}=\frac{1}{P_0+P_1}\Big[&P_0\ket{0000}\bra{0000} + P_1\ket{1111}\bra{1111}\nonumber\\
&+\Gamma\ket{0000}\bra{1111} + \Gamma^*\ket{1111}\bra{0000}\Big],
\label{eqn:ghz-reduced-density}
\end{align}
where
\begin{align}
\Gamma&=\int d\Omega\, \mathcal F(\omega_a,\omega_b,\omega_c,\omega_d)
\mathcal G^*(\omega_a,\omega_b,\omega_c,\omega_d)\nonumber\\
&=\int d\omega_a\,d\omega_b\,d\omega_c\,d\omega_d\,
F_1(\omega_a,\omega_b)F_2(\omega_c,\omega_d)
G_1^*(\omega_a,\omega_c)G_2^*(\omega_b,\omega_d).
\label{eqn:ghz-coherence-general}
\end{align}
We therefore define the normalized GHZ spectral coherence as
\begin{equation}
C_{\rm GHZ}=\frac{|\Gamma|}{\sqrt{P_0P_1}},\qquad 0\leq C_{\rm GHZ}\leq 1.
\label{eqn:normalized-ghz-coherence}
\end{equation}
Its upper bound $C_{\rm GHZ}=1$, following the Cauchy--Schwarz inequality, is reached if and only if the two four-photon spectral amplitudes $\mathcal F$ and $\mathcal G$ are proportional, differing only by a global complex factor.
Equation \eqref{eqn:ghz-coherence-general} and \eqref{eqn:normalized-ghz-coherence} reveal that the GHZ state coherence is not determined by any single photon-pair source alone. 
Instead, it is set by a collective JSA overlap of the four photon-pair sources.

\subsection{State fidelity characterization}

To characterize the experimentally prepared path-encoded states, we measure the fidelity $F = \mathrm{Tr}\left(\ket{\psi}\bra{\psi}\rho_\mathrm{exp}\right)$, where $\ket{\psi}$ is the target state and $\rho_\mathrm{exp}$ is the density matrix of the experimentally prepared state.
For both the EPR and four-photon GHZ states, the target-state projector $\ket{\psi}\bra{\psi}$ can be decomposed into experimentally accessible observables.

For the path-encoded EPR state $\ket{\Phi^+}=(\ket{00}+\ket{11})/\sqrt{2}$, the target-state projector is
\begin{equation}
    \ket{\Phi^+}\bra{\Phi^+} = \frac{1}{4}\left(I\otimes I + Z\otimes Z + X\otimes X - Y\otimes Y\right). 
\end{equation}
Taking its expectation value with $\rho_\mathrm{exp}$ gives the EPR fidelity used in the main text.
For the four-photon GHZ state, the target state is
\[
|G_4\rangle = \frac{1}{\sqrt{2}}\left(|0\rangle^{\otimes4}+|1\rangle^{\otimes4}\right).
\]
Its projector is evaluated using population and coherence observables, giving
\[
F=\frac{1}{2}\langle P\rangle + \frac{1}{2}\langle C\rangle.
\]
The population observable $P=(\ket{0}\bra{0})^{\otimes4}+(\ket{1}\bra{1})^{\otimes4}$ is measured in the computational basis. 
The coherence observable $C$ is obtained from measurements of $M_\theta^{\otimes4}$, where $M_\theta=\cos\theta\,X+\sin\theta\,Y$, as described in the main text.

\subsection{Fourfold counts for four-photon GHZ state fidelity evaluation}

Table \ref{tab:raw-ghz-coincidence} lists the fourfold counts used to extract the four-photon GHZ state fidelity at two EPR state generation probabilities, $p_\mathrm{EPR}$. 
The low-power data are acquired at $p_\mathrm{EPR}=0.012$ with a 10-s integration time per analyzer setting. 
The high-power data are acquired at $p_\mathrm{EPR}=0.026$ with a 3-s integration time per analyzer setting. 
For each probability, the coherence term is measured using four analyzer settings, labelled by the local phases, together with one computational-basis measurement for the population term. 
Each four-character outcome denotes the projection result for photons Q1--Q4, with $+$ or $-$ indicating projection onto the eigenstate with eigenvalue $+1$ or $-1$, respectively.

\begin{table*}[h!]
\caption{
\textbf{Fourfold counts for four-photon GHZ state fidelity evaluation.}
Fourfold counts are listed for the low-power setting $p_\mathrm{EPR}=0.012$ and the high-power setting $p_\mathrm{EPR}=0.026$.
For each $p_\mathrm{EPR}$, the coherence columns correspond to the four analyzer settings labelled by the local phase $\theta$, and the population column corresponds to the computational-basis measurement.
}
\label{tab:raw-ghz-coincidence}
\centering
\small
\setlength{\tabcolsep}{4pt}
\begin{ruledtabular}
\begin{tabular}{lrrrrr@{\hspace{0.35cm}}rrrrr}
\multirow{3}*{Outcome} & \multicolumn{5}{c}{\(p=0.012\), 10 s} & \multicolumn{5}{c}{\(p=0.026\), 3 s} \\
\cline{2-6}\cline{7-11}
 & \multicolumn{4}{c}{Coherence, $\theta$} & \multirow{2}*{Population} & \multicolumn{4}{c}{Coherence, $\theta$} & \multirow{2}*{Population} \\
\cline{2-5}\cline{7-10}
 & 0 & $\pi/4$ & $\pi/2$ & $3\pi/4$ & & 0 & $\pi/4$ & $\pi/2$ & $3\pi/4$ & \\
\colrule
\texttt{++++} & 23 & 0 & 25 & 4 & 162 & 30 & 3 & 41 & 2 & 227 \\
\texttt{+++-} & 0 & 37 & 1 & 25 & 2 & 4 & 47 & 2 & 53 & 4 \\
\texttt{++-+} & 0 & 28 & 3 & 30 & 2 & 0 & 43 & 2 & 64 & 8 \\
\texttt{++--} & 25 & 1 & 27 & 0 & 0 & 31 & 5 & 51 & 3 & 0 \\
\texttt{+-++} & 1 & 41 & 0 & 47 & 3 & 6 & 48 & 10 & 82 & 10 \\
\texttt{+-+-} & 32 & 0 & 42 & 0 & 1 & 44 & 1 & 59 & 5 & 2 \\
\texttt{+--+} & 49 & 7 & 53 & 3 & 1 & 45 & 8 & 68 & 6 & 7 \\
\texttt{+---} & 3 & 43 & 2 & 44 & 1 & 3 & 39 & 6 & 76 & 1 \\
\texttt{-+++} & 2 & 23 & 1 & 20 & 0 & 3 & 25 & 7 & 29 & 5 \\
\texttt{-++-} & 24 & 1 & 16 & 1 & 0 & 26 & 3 & 31 & 1 & 6 \\
\texttt{-+-+} & 19 & 0 & 23 & 0 & 0 & 29 & 0 & 43 & 7 & 6 \\
\texttt{-+--} & 0 & 26 & 1 & 19 & 0 & 1 & 32 & 1 & 50 & 2 \\
\texttt{--++} & 20 & 0 & 29 & 0 & 0 & 28 & 1 & 40 & 3 & 0 \\
\texttt{--+-} & 3 & 28 & 1 & 28 & 0 & 2 & 29 & 4 & 45 & 6 \\
\texttt{---+} & 1 & 37 & 2 & 39 & 0 & 5 & 43 & 1 & 61 & 3 \\
\texttt{----} & 37 & 2 & 36 & 3 & 140 & 27 & 3 & 41 & 11 & 200 \\
\end{tabular}
\end{ruledtabular}
\end{table*}

\section{Fabrication process uniformity}

The film thickness and critical dimension (CD, defined here as the waveguide width) jointly determine the final waveguide geometry and therefore the group index $n_g$. 
For a 200-GHz-FSR microresonator, a 1\% variation in $n_g$ leads to an FSR shift of approximately 2 GHz, much larger than the 100-MHz-level resonance linewidth. 
Therefore, precise resonance alignment across multiple microresonators requires accurate control of $n_g$ and, consequently, stringent fabrication uniformity.

\begin{figure*}[h!]
\centering
\includegraphics[width=0.85\textwidth]{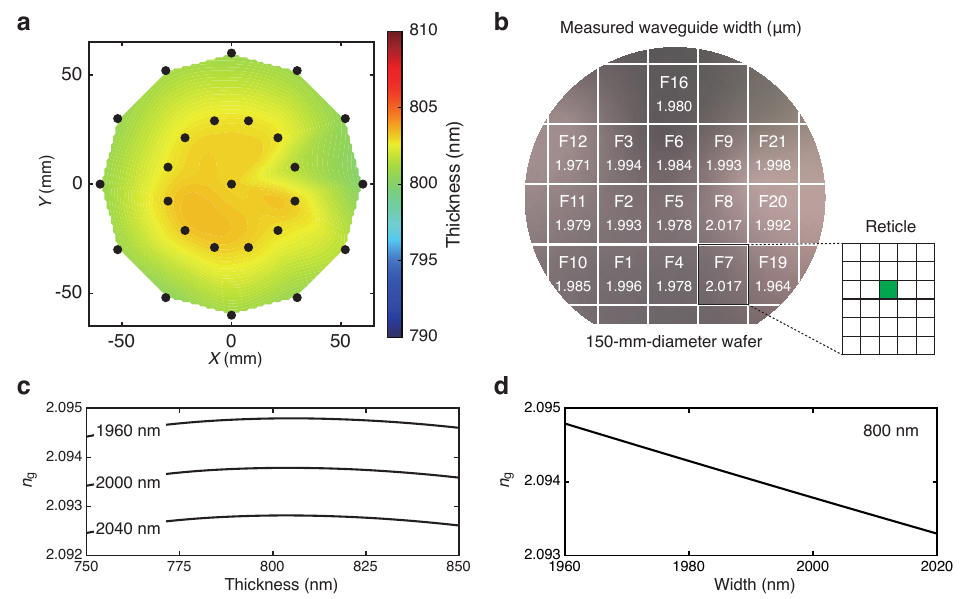}
\caption{
\textbf{Uniformity of the Si$_3$N$_4$ fabrication process.}
\textbf{a}, Measured thickness distribution of the LPCVD Si$_3$N$_4$ film over a $\pm 50$ mm wafer region. 
The film thickness is centered around 803 nm and remains within 800--805 nm over the measured area.
\textbf{b}, Measured critical dimension (CD) of designed 2.000 $\mu$m-wide waveguides across different reticle fields on a 150-mm-diameter wafer. 
The green square in the inset marks the position of the measured test structure within each reticle field.
The numbers indicate the measured waveguide widths in micrometres.
\textbf{c}, Simulated group index $n_g$ as a function of Si$_3$N$_4$ film thickness for waveguide widths of 1960, 2000, and 2040 nm.
\textbf{d}, Simulated group index $n_g$ as a function of waveguide width for a fixed Si$_3$N$_4$ film thickness of 800 nm.
}
\label{Fig:S-uniform}
\end{figure*}

We experimentally evaluate the wafer-level fabrication uniformity. 
Figure~\ref{Fig:S-uniform}a shows the LPCVD silicon nitride film-thickness distribution over a $\pm 50$ mm wafer region. 
The thickness is centered around 803 nm and remains within 800--805 nm. 
We also measure the CD distribution of designed 2.000 $\mu$m-wide waveguides across different reticle fields, as shown in Fig.~\ref{Fig:S-uniform}b. 
After excluding approximately 20\% of fields with large deviations, the CD standard deviation is 0.010 $\mu$m.

As a comparison, Figs.~\ref{Fig:S-uniform}c and \ref{Fig:S-uniform}d show the simulated waveguide group index $n_g$ against waveguide thickness and width, respectively. 
Notably, Fig.~\ref{Fig:S-uniform}c shows that the slope of $n_g$ is close to zero near a thickness of 800 nm, indicating that $n_g$ is insensitive to film-thickness variations in our architecture.
For the waveguide width, a CD variation of 0.010 $\mu$m corresponds to $\Delta n_g \approx 0.00024$, which gives a wafer-level FSR variation of approximately 20 MHz.

\section{Loss budget}

Loss is a key system-level metric for on-chip discrete-variable quantum information processing.
We therefore evaluate the loss budget with the chip for four-photon GHZ state generation, capturing the combined influence of photon extraction from the microresonators, on-chip propagation, chip--fiber coupling, DWDM filtering and detection efficiency.

To measure the end-to-end efficiency, we configure the state-analysis interferometers in the pass-through mode rather than the beam-splitting mode.
Thus photon pairs from the four microresonators are decoupled. 
For each microresonator-based photon-pair source, we measure the two single-photon count rates, $c_1$ and $c_2$, and the corresponding twofold count rate $c_{12}$. 
The measurements are performed at two pump powers used for GHZ state generation, as listed in Table~\ref{tab:loss-budget}. 
Each measurement uses a 0.5-s integration time, and the rates are normalized to hertz.
Background single counts arise from on-chip Raman scattering and residual pump leakage through the DWDM filters. 
To quantify this contribution, we detune the pump away from the microresonator resonances and measure the corresponding single-photon count backgrounds, denoted by $c_{1, \mathrm{bg}}$ and $c_{2, \mathrm{bg}}$. 
The net single-photon count rate is $c_{i,\mathrm{net}}=c_{i}-c_{i, \mathrm{bg}}, ~(i=1,2)$.

\begin{figure*}[b!]
\centering
\includegraphics[width=0.82\textwidth]{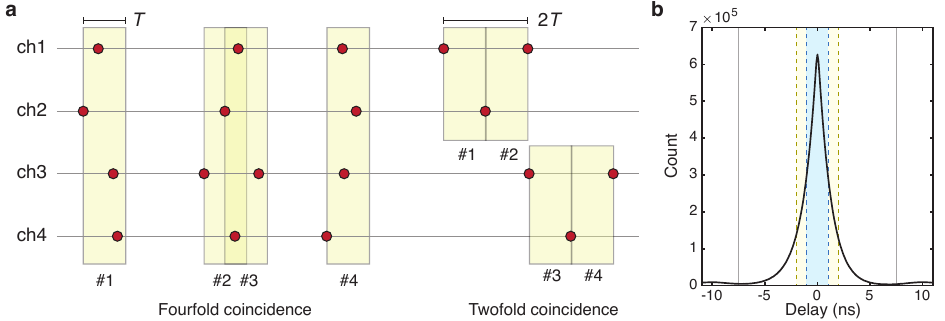}
\caption{
\textbf{Coincidence-window logic used for the loss-budget analysis.}
\textbf{a}, Timing logic for twofold and fourfold coincidence events identification.
The fourfold coincidence events require all four detector clicks to fall within the same coincidence window, while the twofold coincidence events apply the same timing criterion to a pair of detectors.
\textbf{b}, Representative twofold coincidence histogram used to determine the finite-window correction.
The blue shaded region indicates the $\pm$1 ns window used to identify fourfold coincidence events. 
The yellow shaded region indicates the $\pm$2 ns window used to extract $c_{12}$, while the gray vertical lines indicate the $\pm$7.5 ns reference integration range.
}
\label{Fig:S-lossbudget}
\end{figure*}

The coincidence logic must be included when converting the measured count rates into optical efficiencies: all events forming a coincidence must lie within a window of width $T=2$ ns, as shown in Fig. \ref{Fig:S-lossbudget}a.
This window captures only part of the two-photon correlation peak spanning over $\pm 7.5$ ns, as shown in Fig. \ref{Fig:S-lossbudget}b.
The measured twofold count rate is therefore corrected for the finite temporal acceptance of the coincidence logic, which is not associated with optical loss or detector inefficiency, yielding
\begin{equation}
    c_{12,\mathrm{corr}} = 1.23~c_{12}.
\end{equation}
The net single-photon count rate is determined by the photon-pair generation probability $p$ and the overall efficiency $\eta_i$, as
\begin{equation}
    c_{i,\mathrm{net}} = R_\mathrm{rep}\times p\times\eta_i, \quad i=1,2,
\end{equation}
while the corrected twofold count rate is
\begin{equation}
c_{12,\mathrm{corr}} = R_\mathrm{rep}\times p\times\eta_1\eta_2,
\end{equation}
where \(R_\mathrm{rep}=100~\mathrm{MHz}\) is the pump repetition rate.
The measured quantities $c_{1,\mathrm{net}}$, $c_{2,\mathrm{net}}$ and $c_{12,\mathrm{corr}}$ then determine $\eta_1$, $\eta_2$ and $p$ as
\begin{equation}
\eta_1=\frac{c_{12,\mathrm{corr}}}{c_{2,\mathrm{net}}},\quad
\eta_2=\frac{c_{12,\mathrm{corr}}}{c_{1,\mathrm{net}}},\quad
p=\frac{c_{1,\mathrm{net}}c_{2,\mathrm{net}}}{R_\mathrm{rep}c_{12,\mathrm{corr}}}.
\end{equation}

Averaged over the four microresonators, we extract overall efficiencies of $\bar{\eta}=31.2\%$ at high pump power and $\bar{\eta}=30.9\%$ at low pump power.
These values agree with the independently estimated component efficiencies, including
80\% photon extraction efficiency from the microresonator, 
90\% on-chip transmission, 
70\% fiber-array coupling efficiency, 
85\% DWDM filter transmission, 
85\% SNSPD detection efficiency, 
and 85\% transmission through fiber splices and flange connections.
Their product gives $0.80 \times 0.90 \times 0.70 \times 0.85 \times 0.85 \times 0.85\approx 31\%$.
Because each EPR source contains two coherent microresonator excitation alternatives, the EPR state generation probability $p_\mathrm{EPR}$ is twice the mean photon-pair generation probability $\bar p$, $p_\mathrm{EPR}=2\bar{p}$, giving $p_\mathrm{EPR}=0.026$ at the high pump power and $p_\mathrm{EPR}=0.012$ at the low pump power.

\begin{table*}[t!]
\caption{
\textbf{Loss-budget count rates and extracted efficiencies.}
Single-photon count rates $c_1$ and $c_2$, background single-photon count rates $c_{1, \mathrm{bg}}$ and $c_{2, \mathrm{bg}}$, net single-photon count rates $c_{i,\mathrm{net}}=c_{i}-c_{i, \mathrm{bg}}$ and twofold count rates $c_{12}$ are given in hertz.
The corrected twofold count rate is calculated as \(c_{12,\mathrm{corr}}=1.23c_{12}\).
The extracted $p$ is the photon-pair generation probability of one microresonator, and $\eta$ is the end-to-end efficiency.
}
\label{tab:loss-budget}
\centering
\small
\setlength{\tabcolsep}{2pt}
\begin{ruledtabular}
\begin{tabular}{lccccccccccc}
\multirow{2}*{Power} & \multirow{2}*{Source} & \multicolumn{3}{c}{Photon 1} & \multicolumn{3}{c}{Photon 2} & \multicolumn{2}{c}{Twofold} & \multirow{2}*{\(p\)} & \multirow{2}*{\(\eta\)} \\
\cline{3-5}\cline{6-8}\cline{9-10}
 & & \(c_1\) & \(c_{1,\mathrm{bg}}\) & \(c_{1,\mathrm{net}}\) & \(c_2\) & \(c_{2,\mathrm{bg}}\) & \(c_{2,\mathrm{net}}\) & \(c_{12}\) & \(c_{12,\mathrm{corr}}\) & & \\
\colrule
High & 1 & 514178 & 76612 & 437566 & 499308 & 42288 & 457020 & 106490 & 130982.7 & 0.0153 & 29.3\% \\
High & 2 & 528248 & 88536 & 439712 & 401138 & 66944 & 334194 & 94136 & 115787.3 & 0.0127 & 30.2\% \\
High & 3 & 436346 & 110152 & 326194 & 416884 & 54152 & 362732 & 90882 & 111784.9 & 0.0106 & 32.5\% \\
High & 4 & 522070 & 106948 & 415122 & 525324 & 51462 & 473862 & 118680 & 145976.4 & 0.0135 & 32.9\% \\[3 pt]
Low & 1 & 273356 & 82070 & 191286 & 284248 & 53036 & 231212 & 56360 & 69322.8 & 0.0064 & 33.0\% \\
Low & 2 & 297260 & 57322 & 239938 & 234932 & 40590 & 194342 & 51182 & 62953.9 & 0.0074 & 29.2\% \\
Low & 3 & 200850 & 38988 & 161862 & 191542 & 46362 & 145180 & 36088 & 44388.2 & 0.0053 & 29.0\% \\
Low & 4 & 249822 & 69404 & 180418 & 223582 & 56732 & 166850 & 45678 & 56183.9 & 0.0054 & 32.4\% \\
\end{tabular}
\end{ruledtabular}
\end{table*}

This system loss budget further provides an estimate of the fourfold count rate.
For fourfold coincidence events, the $T=2$ ns window is equivalent to integrate the correlation peak over $\pm1$ ns, corresponding to an efficiency factor of 0.57.
The fourfold count rate is thus estimated as
\begin{equation}
    R_{4}=\frac{1}{2}\times R_\mathrm{rep}\times p_\mathrm{EPR}^2\times\bar{\eta}^4\times0.57^2,
\end{equation}
where the factor of $\frac{1}{2}$ is the fusion success probability.
This expression gives $R_{4}=101.4$ Hz for $p_\mathrm{EPR}=0.026$ and $R_{4}=21.6$ Hz for $p_\mathrm{EPR}=0.012$.
These estimates are close to the measured fourfold count rates of 133.7 Hz and 27.0 Hz.

Notably, the system efficiency can be further improved using technologies that are already available.
The photon-extraction efficiency can be increased to approximately 90\% by operating the microresonators in a more strongly over-coupled regime. 
This regime also shortens the photon-pair temporal correlation width, enabling repetition rates of 200 MHz or higher.
Optimized edge-coupler tapers can improve the chip--fiber coupling efficiency to above 80\%. 
In addition, DWDM filters with 90\% transmission can be selected, and SNSPD efficiencies around 90\% are routinely achievable.
The fiber connection losses can be reduced by optimizing the splicing parameters and flange connections, giving an estimated transmission of approximately 95\%.
Combining these improvements yields an expected single-photon end-to-end efficiency of $\sim$50\%, which would substantially increase the detected fourfold count rate to kilohertz level, approaching the performance of state-of-the-art bulk-optical experiments \cite{WangXL:16} while retaining the stability and scalability of the integrated photonic platform.

\vspace{0.2cm}

\clearpage

\section*{Supplementary References}
\bibliographystyle{apsrev4-1}
\bibliography{bibliography, bib_add}